\documentclass[a4paper,11pt]{article}
\usepackage{jheppub} % for details on the use of the package, please see the JINST-author-manual
\usepackage{lineno}
\usepackage{tabularx}
\usepackage{braket}
\usepackage{amsmath}
\usepackage{amsfonts}
\usepackage{amssymb}

\usepackage{xcolor}
\usepackage{graphicx}% Include figure files
\usepackage{dcolumn}% Align table columns on decimal point
\usepackage{bm}% bold math

\usepackage{calrsfs}
\DeclareMathAlphabet{\pazocal}{OMS}{zplm}{m}{n}

\makeatletter
\newcommand*{\bigcdot}{}% Check if undefined
\DeclareRobustCommand*{\bigcdot}{%
  \mathbin{\mathpalette\bigcdot@{}}%
}
\newcommand*{\bigcdot@scalefactor}{.5}
\newcommand*{\bigcdot@widthfactor}{1.15}
\newcommand*{\bigcdot@}[2]{%
  \sbox0{$#1\vcenter{}$}% math axis
  \sbox2{$#1\cdot\m@th$}%
  \hbox to \bigcdot@widthfactor\wd2{%
    \hfil
    \raise\ht0\hbox{%
      \scalebox{\bigcdot@scalefactor}{%
        \lower\ht0\hbox{$#1\bullet\m@th$}%
      }%
    }%
    \hfil
  }%
}
\makeatother

\usepackage{enumitem}

\arxivnumber{2502.11236} % if you have one

\title{\boldmath Weak mixing angle under $\text{U}(1, 3)$ colored gravity}

\author{R. Monjo}
\affiliation{Department of Physics and Mathematics, University of Alcalá \\ Faculty of Sciences, Ctra Madrid-Barcelona 33.6, E-28805 Alcalá de Henares, Madrid, Spain.}
\affiliation{Department of Mathematics, Saint Louis University - Madrid Campus. \\ Manresa Hall, Calle Max Aub, 5, E-28003 Madrid, Spain}
\emailAdd{robert.monjo@slu.edu}

\abstract{Colored gravity, based on $\text{U}(1,3)$ symmetry, emerges naturally in the complexification of Lorentzian manifolds and integrates U(1) electromagnetism as a subcase. This work explores the viability of also including strong and electroweak interactions under the $\text{U}(1,3)$ gauge group of colored gravity. We identify specific generators linked to leptonic and quark interactions and embed the standard Higgs mechanism. Crucially, the weak mixing angle ($\sin^2\theta_W$) is predicted to exhibit about $\sim0.231$ for lepton-lepton interactions (close to observations) and $\sim0.222$ for hadron-lepton interactions, which is in 3$\sigma$ tension with some observations. These findings open pathways for reconciling experimental data with colored gravity and suggest avenues for quantum correction studies.}

\begin{document}
\maketitle
\flushbottom

\section{Introduction}

\subsection{Background of the Weinberg angle}

The weak mixing angle, also known as the Weinberg angle $\theta_W$, is a fundamental parameter in the Standard Model (SM) of particle physics. It quantifies the mixing of the electromagnetic and weak interactions under the electroweak $U(1) \times SU(2)$ gauge symmetry. Expressed as $\sin^2 \theta_W = {q^2}/(q^2 + q'^2)$, where $q$ and $q'$ represent the gauge coupling constants for U(1) and SU(2), respectively, the Weinberg angle governs the masses of the $W$ and $Z$ bosons via $\sin^2 \theta_W = 1 - {m_W^2}/{m_Z^2}$ \cite{Abe2000,Zeller2002}. This parameter plays a pivotal role in precision electroweak physics, offering a bridge between theory and experiment \cite{Abe2000,Zeller2002,CMS_2011,LHCb_2015,CDF_and_D0_2018,LHCb_2024}. While this relationship is theoretically predicted by the SM, the precise value of $\sin^2 \theta_W$ at a given energy scale is influenced by experimentally measured parameters, such as the masses of the W and Z bosons \cite{Ferroglia2013,D0_2015}. An extensive review of low-energy precision measurements for $\sin^2 \theta_W$ was performed in \cite{Kumar2013} for weak neutral-current interactions, mediated by the Z boson.

Precision measurements and renormalization group equations have refined the SM prediction of $\sin^2 \theta_W$ for leptonic interactions, which is approximately $0.23152 \pm 0.00010$ \cite{Ferroglia2013} or $0.23124 \pm 0.00012$ \cite{Davoudiasl2015}. Empirical measurements from experiments such as those carried out at the \textit{SLAC Large Detector} (SLD) of the \textit{Stanford Linear Collider} (SLC), the \textit{Compact Muon Solenoid} (CMS) and \textit{Large Hadron Collider beauty} (LHCb) at CERN, and the \textit{Tevatron collider} at Fermilab have largely corroborated the SM predictions for lepton-lepton interactions (Table \ref{tab:Tab1}), with most results lying within statistical uncertainty \cite{CMS_2011,LHCb_2015,D0_2015,mohr2024}. However, some discrepancies in hadron-lepton processes hint at potential new physics \cite{ALEPH2006}.

\begin{table}
    \setlength\extrarowheight{5pt}
    \caption{Experimental measurement of the effective weak mixing angle $\theta_W$ according to lepton-lepton interactions (upper bock) and for hadron-lepton (lower block).}
    \label{tab:Tab1}
\begin{tabular}{p{3.72cm} p{2.41cm} p{2.82cm} p{3.82cm}} %{tabularx}{\textwidth}{XXXX}%{lrll}
\hline
\hline
Source & $\sin^2\theta_W$ & Uncertainty & Reference \\
\hline
SLD & 0.23098 & $\pm$ 0.00026 & SLD Col. 2000 \cite{Abe2000} \\
NuTeV & 0.2277 & $\pm$ 0.0016 & Zeller+. 2002 \cite{Zeller2002} \\
CMS/LHC & 0.2287 & $\pm$ 0.0020 (stat) \; $\pm$ 0.0025 (syst) & CMS Col. 2011 \cite{CMS_2011} \\
LHCb/LHC & 0.23142 & $\pm$ 0.00052 (stat) $\pm$ 0.00056 (syst) & LHCb Col. 2015 \cite{LHCb_2015} \\
Low $Q^2$ & 0.2328 & $\pm$ 0.0009& Davoudiasl+ 2015 \cite{Davoudiasl2015} \\
D0/Tevatron & 0.23147 & $\pm$ 0.00047 & D0 Col. 2015 \cite{D0_2015} \\
D0/Tevatron & 0.23095 & $\pm$ 0.00035 (stat) $\pm$ 0.00020 (syst) & CDF+D0 Col. 2018 \cite{CDF_and_D0_2018} \\
CDF+D0 (Tevatron) & 0.23148 & $\pm$ 0.00033 & CDF+D0 Col. 2018 \cite{CDF_and_D0_2018} \\
LHCb/LHC & 0.23147 & $\pm$ 0.00044 (stat) $\pm$ 0.00005 (syst) & LHCb Col. 2024 \cite{LHCb_2024} \\
\hline
D0/Tevatron & 0.22269 & $\pm$ 0.00034 (stat) $\pm$ 0.00021 (syst) & CDF+D0 Col. 2018 \cite{CDF_and_D0_2018}
\\
CDF+D0 (Tevatron) & 0.22324 & $\pm$ 0.00033
& CDF+D0 Col. 2018 \cite{CDF_and_D0_2018}
\\
CODATA & 0.22305 & $\pm$ 0.00023 & Mohr+ 2024 \cite{mohr2024} \\
\hline
\hline
\end{tabular}
\end{table}
%Mohr, P. J., Taylor, B. N., & Newell, D. B. (2024). "CODATA recommended values of the fundamental physical constants: 2022. ArXiv https://doi.org/10.48550/arXiv.2409.03787

Despite the success of the electroweak framework, integrating quantum chromodynamics (QCD) into a unified theory remains a profound challenge. Grand unified theories (GUTs) aim to merge the electroweak and strong interactions into a single theoretical framework \footnote{A GUT proposal must have at least rank 4. Recall that the rank of a Lie group is equal to the number of diagonal (mutually commuting) generators in its Cartan subalgebra. For example, the $\text{rank}\,SU(1,p) = \text{rank}\,SU(1+p) = p$, while the factor $U(1)$ embedded in $\text{U}(1,3)$ provides an additional diagonal matrix (the unitary), so finally the rang of $\text{U}(1,3)$ is 4.}, potentially refining predictions for $\sin^2 \theta_W$ \cite{Singer1982,Oshimo2009,Senjanovi2024,Basiouris2025}.  For instance, the SU(5) GUT predicts a value of $\sin^2\theta_W = 3/8 \approx 0.375$ at the unification energy scale, which is around $10^{16}$ GeV \cite{Dimopoulos2002, Babu2024}. When running the renormalization group equations down to the electroweak scale (around the Z boson mass), the predicted value of $\sin^2\theta_W \approx 0.21$ \cite{Einhorn1982,Dimopoulos2002,Kumar2013}. Even smaller is the mixing angle for the $\text{SU}(5) \times \text{SU}(5)$ theory (or double $\text{SU}(5)$), with $\sin^2\theta_W = 3/16 = 0.1875$ at a unification energy scale; although it is expected to increase at lower energy scales \cite{Babu2024}. Similarly, a minimally supersymmetric SO(10) can obtain values about $\sin^2\theta_W \approx 0.2210$ \cite{Aulakh2012}.

\subsection{Motivation and objectives}

\subsubsection{Motivation for $U(1,3)$-based colored gravity}

Double copy of $\mathfrak{su}(N)$ and more generally double $\mathfrak{u}(N)$ for $N=4,5$ contain relevant subalgebras that are related to the lepton-quark interactions \cite{RedKov2008, Castro2012, CembranosDiezValle2019}. The $\mathfrak{su}(4)$ algebra provides a framework to unify quarks and leptons within the same symmetry group \cite{MarschandNarita2015, BarbieriandTesi2018}, whereas $\mathfrak{u}(1,3)$ plays a significant role in describing the strong interactions of quarks and gluons in the non-perturbative regime at large interaction distances \cite{Khruschev2004}. Within this context, it is plausible that the $\mathfrak{u}(1,3)$ algebra could also support a quark-lepton unification model, analogous to the $\mathfrak{su}(4)$ case. Furthermore, $\mathfrak{u}(1,3)$ exhibits connections to Wess-Zumino-Witten models in two dimensions and Chern-Simons theories in three dimensions \cite{Margolin1992, Tseytfin1995}.

More recently, a proposal of colored gravity was successfully based on a double $\mathfrak{su}(1,3)$ subalgebra \cite{Monjo2024}. Specifically, using a gauge-like treatment of \textit{teleparallel gravity equivalent to general relativity} (TEGR), Lagrangian density of colored gravity is isomorphic to a $\text{SU}(1,3)$ Yang--Mills (YM) theory \cite{deAndradePereira1997, Krssak2019, Monjo2024}.

Colored gravity is motivated by the idea that $\text{U}(1,3)$ gauge group emerges from the complexification of Lorentzian manifolds and spinor field dynamics, offering a natural extension to unify gravity with the SM \cite{Monjo2024}. In this framework, $\text{U}(1)$ electromagnetism appears as a subset, while $\text{SU}(3) \subset \text{U}(1,3)$ represents the strong interaction. The structure also supports a Higgs mechanism that integrates leptonic and quark interactions into a unified description. 

%Recent studies have shown that this approach yields predictions for $\sin^2 \theta_W$ consistent with empirical measurements for lepton-lepton processes, while highlighting potential discrepancies in hadron-lepton interactions.

\subsubsection{Objectives and structure of this work}

This paper aims to extend the SM by embedding its symmetries into the $\text{U}(1,3)$ gauge group, providing insights into the weak mixing angle and its implications for unification. The key contributions include:

\begin{itemize}[noitemsep, topsep=0pt]
    \item Showing a natural embedding of $\text{SU}(3)$ and $\text{U}(1) \times \text{SU}(2)$ symmetries into the $\text{U}(1,3)$ framework.
    \item Extending the Higgs mechanism to $\text{U}(1,3)$, linking the electroweak symmetry-breaking process to broader unification.
    \item Providing theoretical predictions for the weak mixing angle under different interaction contexts.
\end{itemize}

The paper is structured in five main sections. After the preliminary Sec. \ref{sec:prel} aimed to set the foundational notation, Sec. \ref{sec:colored} introduces the key features of $U(1,3)$-based colored gravity. Sec. \ref{sec:embedding_SM} describes the embedding of the SM algebra representatives within $U(1,3)$ and identifies each fundamental interaction. Then Sec. \ref{sec:higgs_mechanism} develops the fit of Higgs mechanism into the $\text{U}(1,3)$ model by identifying the generators involved and provides a first prediction of the weak mixing angle. Finally, Sec. \ref{sec:conclusions} collects the main insights and concluding remarks to interpret the results and outlines directions for future research.

\section{Preliminaries}
\label{sec:prel}

\subsection{Spacetime algebra}
\label{sec:spacetime_algebra}

This subsection summarizes the key concepts from \cite{Gu2018, Gu2018b}. Let $(M, \mathbf{g})$ be a 4-dimensional Lorentzian manifold with Minkowski bundle $\mathcal{M} \to \mathrm{T}M$ and frames $\{x^\mu, x^a\}_{\mu,a}$. The spacetime algebra $Cl_{1,3}(\mathbb{R}, \mathbf{h})$, with $\mathbf{h} = \eta$ or $\mathbf{g}$, is generated by $\{v_\mu\}_\mu$ and $\{\gamma_a\}_a$, which satisfy:
\begin{align}
v_\mu \bigcdot v_\nu &= \tfrac{1}{2}(v_\mu v_\nu + v_\nu v_\mu) = g_{\mu\nu}\mathbf{1}_4, \\
\gamma_a \bigcdot \gamma_b &= \tfrac{1}{2}(\gamma_a \gamma_b + \gamma_b \gamma_a) = \eta_{ab}\mathbf{1}_4,
\end{align}
where $\bigcdot$ is the symmetric (dot) product, and $\mathbf{1}_4$ is the identity matrix. The antisymmetric (wedge) product is defined as:
\begin{align}
v_\mu \wedge v_\nu &= \tfrac{1}{2}(v_\mu v_\nu - v_\nu v_\mu) = \tfrac{1}{2}[v_\mu, v_\nu], \\
\gamma_a \wedge \gamma_b &= \tfrac{1}{2}(\gamma_a \gamma_b - \gamma_b \gamma_a) = \tfrac{1}{2}[\gamma_a, \gamma_b].
\end{align}

The algebra generators decompose into symmetric and antisymmetric parts: $v_\mu v_\nu = v_\mu \bigcdot v_\nu + v_\mu \wedge v_\nu$, and similarly for $\gamma_a \gamma_b$. The tetrad components relate to spacetime generators via $v^\mu = e^\mu_{\ a}\gamma^a$ and $\gamma^a = e_\mu^{\ a}v^\mu$, satisfying orthogonality: $\gamma_a \bigcdot \gamma^b = \delta_a^b \mathbf{1}_4$ and $v_\mu \bigcdot v^\mu = \gamma_a \bigcdot \gamma^a = \mathbf{1}_4$.

The symbols $v_\mu$ and $\gamma_a$ emphasize their connections to the four-velocity covector $u_\mu = dx_\mu/d\tau$ and Dirac gamma matrices, respectively. Distances defined by the generators mirror those from $u_\mu dx^\mu$. The spacetime element is $\mathbf{d\boldsymbol{\tau}} = v_\mu dx^\mu = \gamma_a dx^a$, and the metric is derived as:
\begin{align*}
\mathbf{1}_4 d\tau^2 = \mathbf{d\boldsymbol{\tau}} \bigcdot \mathbf{d\boldsymbol{\tau}} &= (v_\mu \bigcdot v_\nu) dx^\mu dx^\nu = \mathbf{1}_4 g_{\mu\nu} dx^\mu dx^\nu \\
&= (\gamma_a \bigcdot \gamma_b) dx^a dx^b = \mathbf{1}_4 \eta_{ab} dx^a dx^b.
\end{align*}
This formalism encapsulates spacetime geometry in terms of Clifford algebra, providing a robust foundation for describing both the kinematics and dynamics of fields in curved spacetimes. By leveraging these structures, our framework enables an extension of SM symmetries to incorporate gravity through the \textcolor{black}{deformation} of the spacetime generators by the $\mathfrak{u}(1,3)$ algebra\textcolor{black}{, which may imply a noncommutative spacetime structure. This possibility, and its implications for Lorentz invariance, are further discussed in Sec. \ref{sec:impact_lorentz}}

%For a vector field $w^a$, the total time derivative is $\mathbf{D_\tau}w^a := (v^\mu D_\mu)w^a = \mathbf{d_\tau}w^a + \eta_{bc}\Gamma_\mu^{\ ab}w^c v^\mu$, where $D_\mu$ is the Fock-Ivanenko covariant derivative and $\mathbf{d_\tau} = v^\mu \partial_\mu = \gamma^a \partial_a$. %%The classical definition arises by replacing $v^\mu \mapsto u^\mu$ and using the standard product.

\subsection{Complexified Minkowskian spacetime and spinors}

Let $(\mathcal{M}, \eta)$ represent the Minkowskian manifold with the metric $\eta = \mathrm{diag}(1, -1, -1, -1)$, and let $(\mathcal{M}^c, \eta^c)$ denote its \textit{complexification}, defined as $\mathcal{M}^c := \mathcal{M} \oplus \mathbf{i}\,\mathcal{M}$, where $\mathbf{i} := e_0 e_1 e_2 e_3$ is the unit pseudoscalar constructed using the vector basis $\{e_\mu\}_{\mu=0}^4$ of $\mathcal{M}$ \cite{Friedman2021,Chappell2023}. The complexification introduces additional degrees of freedom, allowing the description of both \textit{real} (e.g., velocity and momentum) and \textit{imaginary} components (e.g., angular momentum and magnetic moment), whose behavior under parity inversion differs. The terms \textit{imaginary} and \textit{complexification} arise from the property $\mathbf{i}^2 = (e_0 e_1 e_2 e_3)(e_0 e_1 e_2 e_3) = -1$, with $e_i e_j := e_i \bigcdot e_j + e_i \wedge e_j$.

In the framework of $\mathcal{M}^c \ni \mathrm{a}, \mathrm{b}$, the inner product structure is defined by
\begin{eqnarray}\nonumber
    \eta(\mathrm{a}, \mathrm{b}) & := &\eta(a+\mathbf{i}\,x,\;  b + \mathbf{i}\,y) := \\ \nonumber
    &:=& \eta(a,b)+\eta(x,y)-\mathbf{i}\,\eta(x,b) + \mathbf{i}\,\eta(a,y)\,,
\end{eqnarray}
where $\mathrm{a} = a+\mathbf{i}\,x$ and $\mathrm{b} = b + \mathbf{i}\,y$, with $a$, $b$, $x$, and $y$ being real tangent (vector) fields on $\mathcal{M}$. This definition ensures that both the real and imaginary components of vectors contribute to the overall geometry in a consistent manner. 

As an example, any spinor (a spin-$\frac{1}{2}$ particle) can be expressed as $\psi = u + \mathbf{i}\,s \in \mathcal{M}^c$, where $u$ represents the unit four-velocity or its \textit{linear momentum}, and $s$ denotes the four-dimensional spin angular momentum, which is linked to the Pauli-Lubanski pseudovector \cite{Chappell2023}.  \textcolor{black}{The spacetime coordinates are therefore formulated as spinor-valued entities within the complexified 
$Cl_{1,3}(\mathbb{R}) \otimes \mathbb{C}$ algebra. This representation explicitly extends the real Lorentzian spacetime to a complex manifold $\mathcal{M}^c$, enabling a unified description of momentum and spin dynamics while preserving Lorentz invariance, that is, into} a relativistic two-state prescription. For example, the interaction of an electron (or similar charged particle) with mass $m$, spin $s$, and Landé factor $\mathrm{g_e} \approx 2$, in the presence of an electromagnetic field $F$, can be described as:
\begin{align}
\frac{d}{d\tau}\psi^\alpha \approx \mu \psi^\beta F_{\beta}^{\ \alpha},
\end{align}
where $\mu = \frac{\mathrm{g}_e}{2}\frac{q}{m}s \approx \frac{q}{m}s$ is the magnetic moment \cite{Friedman2021}. This equation highlights the coupling between the spinor's intrinsic properties and the external field.

In the Dirac spinor formalism, both the electron (particle) and positron (antiparticle) are represented within the four-component spinor 
$\psi = (\chi\; \chi^c)^\top$, satisfying the Dirac equation 
$(i \gamma^\mu \partial_\mu - m)\psi = 0$. The upper two components (represented by $\chi$ correspond to the electron, while the lower components (from $\chi^c$) describe the positron, whose interpretation as an antiparticle arises via charge conjugation: 
$\psi^c = C \overline{\psi}^\top$, where $C$ is the charge conjugation matrix, $\overline{\psi}$ is the complex conjugate, and therefore $(\chi^c)^c = \chi$. This unified description captures both particle and antiparticle dynamics within a single framework.
 
The four-component spinor $\psi$ can be expressed in the Weyl or chiral representation as $\psi_\text{chiral} = \psi_L + \psi_R = (\chi_L\; 0)^\top + (0\; , \chi_R)^\top = (\chi_L\; \chi_R)^\top$, where $\chi_L = \tfrac{1}{\sqrt{2}}\left(\chi + \chi^c\right)$ and $\chi_R= \tfrac{1}{\sqrt{2}}\left(\chi - \chi^c\right)$ are the two-component Weyl spinors in the same chiral basis\footnote{This is equivalent to apply the unitary matrix $U_{chiral} = \tfrac{1}{2}\left(\begin{matrix}
    I & I \\ I & -I
\end{matrix}\right)$ to $\psi \mapsto \psi_\text{chiral} = U_{chiral}\psi$, and therefore $\psi_L = P_L U_{chiral}\psi$ and $\psi_R = P_R U_{chiral}\psi$}. Then, using the Clifford algebra $\{\gamma^i\}_{i=0}^3$ for the left- and right-hand projectors $P_L = \frac{1}{2}(1 - \gamma^5)$, $P_R = \frac{1}{2}(1 + \gamma^5)$ with $\gamma^5 := i\gamma^0\gamma^1\gamma^2\gamma^3$, the spinor can be decomposed as $\psi_L = P_L\psi_\text{chiral}$ and $\psi_R = P_R\psi_\text{chiral}$. Moreover, the Dirac equation is now $(E - \mathbf{\sigma} \cdot \mathbf{p}) \psi_R = m \psi_L$ and $(E + \mathbf{\sigma} \cdot \mathbf{p}) \psi_L = m \psi_R$ with Pauli matrices $\mathbf{\sigma}$, momentum $\mathbf{p} = m\mathbf{v} \in \mathbb{R}^3$ and energy $E^2 := m^2 + \mathbf{p}^2$. In contrast to the Dirac representation, this decomposition enables a detailed analysis of left- and right-handed states, which are fundamental to weak interactions.

%https://arxiv.org/abs/1908.04590
%https://arxiv.org/pdf/1908.04590
%https://www.scirp.org/pdf/JMP_2015011416452320.pdf

In the context of particle dynamics, the four-velocity is $u^\mu = \tfrac{1}{\sqrt{1-\mathbf{v}^2}}(1, \mathbf{v})$, and the spin pseudovector is 
$s^\mu = \frac{1}{2} \epsilon^{\mu\nu\rho\sigma} u_\nu J_{\rho\sigma}$, describing the motion and spin of both particle and antiparticle. For an electron at rest, 
$u^\mu = (1, 0, 0, 0)$ and $s^\mu = (0, 0, 0, \pm\frac{1}{2})$, where $\pm\frac{1}{2}$ denotes spin alignment. The positron's $u$ and $s$ have the same forms but correspond to its specific quantum state, with negative energy solutions of the Dirac equation reinterpreted as positive via field quantization.

To robustly compute these quantities from a spinor $\psi$, bilinear covariants are obtained from $u^\mu = \overline{\psi} \gamma^\mu \psi$ and $s^\mu = \frac{1}{2} \epsilon^{\mu\nu\rho\sigma} \overline{\psi} \gamma_\nu \gamma_{\rho\sigma} \psi$, where $\overline{\psi} = \psi^\dagger \gamma^0$ is the Dirac adjoint, while $\epsilon^{\mu\nu\rho\sigma}$ is the Levi-Civita symbol and $\gamma_{\rho\sigma}$ are the Lorentz transformation generators $\gamma_{\rho\sigma} = \frac{1}{2} [\gamma_\rho, \gamma_\sigma]$. These expressions ensure that $u^\mu$ and $s^\mu$ are well-defined geometric quantities\footnote{\textcolor{black}{These quantities arise from bilinear covariants constructed with Clifford algebra, providing a coordinate-free geometric interpretation.}} derived from the spinor $\psi$ and the Clifford algebra structure, making them representation-independent. For instance, the four-velocity from the chiral representation is $u^0 = \chi_L^\dagger \chi_R + \chi_R^\dagger \chi_L = 1,\;
u^i = \chi_L^\dagger \sigma^i \chi_R - \chi_R^\dagger \sigma^i \chi_L = 0$ for a spin up particle at rest $\chi_L = \chi_R = \tfrac{1}{\sqrt{2}}(1\; 0)^\top$, while the spin pseudo-vector is $s^\mu = (0,0,0,+\tfrac{1}{2})$.

%In the following, all the notations previously introduced are assumed. 

\subsection{Natural U(1, 3) group in gravity}
\label{sec:natural}

\noindent Let $(M, g)$ represent an oriented Lorentzian 4-manifold with metric signature $(+, -, -, -)$, equipped with a spinor structure or an oriented loop space $\pazocal{L}M$.

The real tangent bundle of $M$, denoted by $\mathrm{T}M$, can be complexified to $\mathrm{T}^cM = \mathrm{T}M \otimes \mathbb{C} \cong \mathcal{M}^c$. Its associated frame bundle forms a principal $\mathrm{GL}(4, \mathbb{C})$-bundle, written as $L^c(M) \to M$. The real Lorentzian metric $g$ on $\mathrm{T}M$ extends naturally to $\mathrm{T}^cM$ as a Hermitian form:
\begin{equation*}
    g(a+\mathbf{i}x, b+\mathbf{i}y) = g(a, b) + g(x, y) - \mathbf{i} g(x, b) + \mathbf{i} g(a, y),
\end{equation*}
where $a, b, x, y$ are real vector fields on $M$. This extension preserves the Lorentzian signature $(+, -, -, -)$ while introducing a consistent framework for handling complexified spacetime geometries.

Reducing $L^c(M) \to M$ to a principal $\mathrm{U}(1,3)$-bundle requires preserving the Hermitian structure of the metric. Additionally, the volume form defined by $g$ and the orientation of $M$ reduces the structure group further to $\mathrm{SL}(4, \mathbb{C})$. Combining these reductions, one obtains a principal bundle with structure group $\mathrm{U}(1,3) \cap \mathrm{SL}(4, \mathbb{C}) = \mathrm{SU}(1,3)$ \footnote{According to \cite[Proposition 5.6]{kobayashinomizu1963}, such a reduction corresponds to a global section of the bundle of oriented orthonormal frames over $M$. Formally, this section exists if and only if
\begin{equation*}
\frac{L^c(M) \times \mathrm{GL}(4, \mathbb{C}) / \mathrm{SU}(1,3)}{\mathrm{GL}(4, \mathbb{C})} \to M,
\end{equation*}
where $\mathrm{GL}(4, \mathbb{C})$ acts on $\mathrm{GL}(4, \mathbb{C}) / \mathrm{SU}(1,3)$ via left multiplication.}.

The Levi--Civita connection $\overset{\circ}{D}$ induced by $g$ extends to $\mathrm{T}^cM$ as $\overset{\circ}{D}{}^c$, serving as a torsion-free reference connection among all possible $\mathrm{SU}(1,3)$ connections on $\mathrm{T}^cM$. This torsion-free extension ensures compatibility between the classical geometry of $M$ and the complexified spacetime.

Within this framework, the gauge theory of the non-compact group $\mathrm{U}(1,3)$ is considered, a topic that has long been debated in relation to the properties of quantum field theories (QFTs) based on such gauge groups \cite{Tseytfin1995,weinberg2005,Fabbrichesi2021}. Managing the non-compact nature of these groups often requires gauge-fixing methods, such as the Faddeev-Popov procedure, where ghost fields and techniques like the Lorenz gauge ensure unitarity and identify the physical degrees of freedom \cite{Thierry1980,Eichhorn2013}. While addressing the complexities of quantization is outside the scope of this work, we assume a framework bridging the classical $\mathrm{SU}(1,3)$ YM formalism with a gauge-like treatment of teleparallel gravity, as described in \cite{Monjo2024}. The analogy to chromodynamics motivates the term \textit{colored gravity}, emphasizing its connection to the \textit{strong force}.

For simplicity, we take $M \cong \mathcal{M}$ to be Minkowski spacetime, which is a parallelizable and non-compact manifold. This allows for a global section of the frame bundle, thereby admitting a spinor structure. The spinor structure is essential for defining a $\mathrm{U}(1,3)$-colored connection, which governs the parallel transport and covariant differentiation of spinor fields. At this stage, perturbations of the Minkowski metric involving spinors are considered negligible for the purpose of defining the spin structure.

\section{Colored gravity}
\label{sec:colored}

\subsection{A gauge-like treatment of gravity}

The colored gravity theory proposed by \cite{Monjo2024} is based on the idea of a classical-to-quantum bridge between the SU$(1, 3)$ YM gauge formalism and the gauge-like treatment of teleparallel gravity. This framework provides a novel unification by \textcolor{black}{linking an extended TEGR to} a $\mathrm{U}(1,3)\times \mathrm{U}(1,3)$ gauge theory, offering a reinterpretation of gravity as a gauge-like interaction \textcolor{black}{(see details in Appendix~\ref{sec:basics})}. 

Particularly, spacetime algebra \textcolor{black}{extended} with \textcolor{black}{the} Weitzenböck connection can be assimilated to a local complexification based on the SU$(1,3)$ YM theory producing Maxwell-like dynamics \cite{deAndradePereira1997,Itin2006,Krssak2019}. As mentioned in Sec. \ref{sec:natural}, the pseudo-unitary group $\text{U}(1,3)$ is naturally found in the complex hyperbolic space $\mathbb{CH}^3$, also noted as $\mathbb{H}_{\mathbb{C}}^3$. This space, characterized as a Kähler manifold \cite{Pozetti2014,DiazRamos2023}, possesses three mutually compatible structures: a complex structure, a Riemannian structure, and a symplectic structure, making it a versatile mathematical framework for exploring unified theories. In colored gravity, YM dynamics emerges from locally-\textcolor{black}{deformed} tetrads, recovering features of classical electrodynamics \citep{Monjo2024}.  \textcolor{black}{The term `deformation' here refers to how the U(1,3) gauge potential $A_\mu$ extends the spacetime algebra, modifying both spinor frames and local geometric structures. While formally analogous to YM fluctuations ---where the background remains Minkowskian but acquires gauge corrections \citep{Weinberg1996,Ferko2025}--- this framework extends beyond small deviations to encompass finite deformations of the algebraic structure under $\mathfrak{u}(1,3)$ symmetry.}

%(Sec.~\ref{sec:perturbedspinor}) 

%In our framework, U(1,3) gauge fluctuations modify the spacetime algebra itself (Eq.~3.3), unlike Yang--Mills theories where fluctuations occur on a fixed background. This aligns with teleparallel gravity, where the gauge connection directly determines the spacetime geometry through torsion.

%The colored gravity hypothesis \cite{Monjo2024} posits that the Cartan covariant derivative for gravity resembles an $\mathrm{SU}(1,p)$ or $\mathrm{U}(1,p)$ gauge derivative for $p = 3$. Here, spacetime generators $\gamma_a$ are perturbed by a $\mathrm{U}(1,p)$ gauge potential $A_\mu$, involving the matter-energy gauge phase. This leads to $v^\mu \approx e^\mu_{\ a}\gamma^a + O(A_\mu)$, suggesting that $\{v_\mu\}_\mu$ encodes dynamics related to the Standard Model.

\subsection{\textcolor{black}{Transformed} spinors}
\label{sec:perturbedspinor}

Let $\Psi = \{\psi_n\}_{n=1}^4$ be a multiplet of four Dirac spinors on $M$, where each spinor $\psi_n$ consists of four components and represents fermionic fields. These spinors can be written as $\ket{\Psi} \in \pazocal{M}_4(\mathbb{C}) = \mathbb{C} \times Cl_{1,3}(\mathbb{R}, g)$. The total spin of $\Psi$ ranges from $-2$ to $2$, and the spinor fields describe particles with the same energy $m > 0$, encoded in the four-momentum $\mathbf{m}$, satisfying $\mathbf{m} \bigcdot \mathbf{m} = m^2 \mathbf{1}_4$. The components of $\mathbf{m}$ are given by $m_\mu = \mathbf{m} \bigcdot \gamma_\mu$, linked to the classical momenta by $m_\mu = m u_\mu \mathbf{1}_4 := m \frac{dx_\mu}{d\tau} \mathbf{1}_4$. Consequently, the velocity is $\mathbf{u} := \mathbf{m}/m = \gamma^\mu u_\mu$, and the spacetime element is $\mathbf{d}\boldsymbol{\tau} := \gamma_\mu dx^\mu$, leading to \footnote{We introduce the spacetime dependence of a spinor field as a phase-displacement operator $\exp\left(-{{i}}\, \mathbf{m} \bigcdot \mathbf{d}\boldsymbol{\tau}\right)$ applied to the initial phase $\Psi_0(\mathbf{m})$.}:
\begin{equation}
\Psi = \exp\left(-{{i}}\, \mathbf{m} \bigcdot \mathbf{d}\boldsymbol{\tau}\right) \Psi_0(\mathbf{m}).
\end{equation} This phase formalism elegantly links the momentum of spinor fields to spacetime evolution.

The gauge potential $\mathbf{A} = A_\mu \gamma^\mu$ is also expressed via the spacetime algebra \cite{Dressel2015}, where $A = A_\mu dx^\mu \in \Omega^1(M, \mathfrak{u}(1,3))$ represents a connection on the $\text{U}(1, 3)$-bundle with components $A_\mu = A_\mu^I \ell_I$ in a basis $\{\ell_I\}_{I=0}^{15}$ of $\mathfrak{u}(1,3)$. \textcolor{black}{The gauge fields are assumed to transform under the adjoint representation of U(1, 3), in line with standard treatments of gauge fields in YM theories (Appendix~\ref{sec:u13} and \ref{sec:adjoint_representation})}. The origin $\mathbf{A}(0)$ of the U(1, 3) gauge potential can be defined using covariant Liénard-Wiechert or Cornell-like forms. Normalized using the quadratic Casimir operator, we will choose an origin $\mathbf{A}(0)$ proportional to the momentum $\mathbf{m} = m^\mu \gamma_\mu = m_\mu \gamma^\mu$ as follows:
\begin{equation*}
\mathbf{A}(0) \equiv \frac{q}{\kappa \mathbf{m}} = \frac{q}{\kappa m^2} \, \mathbf{m} \; \in \mathcal{M}_4(\mathbb{C}),
\end{equation*} where $q$ is the $\text{U}(1,3)$ coupling constant (or from the subgroup considered) and $\kappa = 8\pi\mathrm{G}$ is the gravitational constant, built with the Newtonian constant $\text{G}$.

The associated covariant derivative transforms as $\nabla_\mu = \partial_\mu - {{i}}q A_\mu$, introducing a phase $\varphi(x) = -q A_\mu dx^\mu$ into the spinor field. This yields the unitary transformation $U(x) = \exp({{i}}\varphi(x))$ and the transformed spinor
\begin{equation}
\label{eq:transformation_spinor}
\Psi \mapsto \hat{\Psi} := U(x) \Psi =  \exp\left(-{{i}} \left(\mathbf{m} \bigcdot \gamma_\mu + q A_\mu\right) dx^\mu\right)\Psi_0(\mathbf{m}).
\end{equation}
The covariant derivative $\nabla_\mu$ remains symmetric under the transformation, $\hat{\Psi}^\dagger \hat{\nabla}_\mu \hat{\Psi} = \Psi^\dagger \nabla_\mu \Psi$, where $\hat{\nabla}_\mu = U(x) \nabla_\mu U^\dagger(x)$. The unitary operator $U^\dagger(x)$ is connected to the Wilson loop and generalizes the Aharanov-Bohm effect \cite{Alfonsi2020}.

The compensatory phase $\varphi(x)$ modifies the spacetime generators $\gamma_\mu$, resulting in the \textcolor{black}{deformed} generators
\begin{equation}
\label{eq:gamma_matrix2}
\gamma_\mu \mapsto \hat{\gamma}_\mu := \gamma_\mu + \kappa \, A_\mu \bigcdot \mathbf{A}(0) =  \gamma_\mu + A_\mu \bigcdot \frac{q}{\mathbf{m}},
\end{equation}where $1/\mathbf{m} := \mathbf{m}/m^2$. Similarly, the momentum $\mathbf{m}$ and velocity $\mathbf{u}$ are \textcolor{black}{transformed} as
\[
\mathbf{m} \mapsto \hat{\mathbf{m}} := \mathbf{m} + q\mathbf{A}, \quad \mathbf{u} \mapsto \hat{\mathbf{u}} := \frac{\hat{\mathbf{m}}}{m} = \mathbf{u} + \frac{q \mathbf{A}}{m}.
\]
Thus, the \textcolor{black}{transformed} velocity satisfies
\begin{equation}
u_\mu \mapsto \hat{u}_\mu := \mathbf{u} \bigcdot \hat{\gamma}_\mu = \mathbf{1}_4 u_\mu + A_\mu \frac{q}{m},
\end{equation}
where $\hat{u}_\mu \in \mathbb{R} \oplus \mathfrak{u}(1,3)/\mathbb{R} = \mathbb{R} \oplus \mathfrak{su}(1,3)$ \footnote{Here, $\mathbb R$ is regarded as a subspace of $\mathfrak{gl}(4,\mathbb C)$ via the injection $a\mapsto a\mathbf 1_4$.}. This demonstrates how spacetime translations, such as $\delta \hat{\mathbf{x}} = \hat{\mathbf{u}} \, \delta \tau = \gamma^\mu \hat{u}_\mu \delta \tau$, extend naturally within the new framework, generalizing the Poincaré algebra\footnote{\textcolor{black}{Although the spacetime coordinates are formally expressed using spinors, the formulation retains a classical Lorentz-invariant structure via Clifford algebra.}}. These constructions provide a geometric foundation for describing \textcolor{black}{transformed} spinor dynamics in curved or complexified spacetime geometries. 

\textcolor{black}{In our framework, U(1,3) gauge fluctuations modify the spacetime algebra itself (Eq.~\ref{eq:gamma_matrix2}), unlike YM theories, where fluctuations occur on a fixed background. This aligns with teleparallel gravity, where the gauge connection directly determines the spacetime geometry through torsion.}

%$(\mathbb{R}^{1,3} \oplus \mathfrak{u}(1,3)) \otimes (\mathbb{R}^{1,3} \oplus \mathfrak{u}(1,3))$

%This construction demonstrates how translations $\delta \hat{\mathbf{x}} = \hat{\mathbf{u}} \, \delta \tau$ lie in $\mathbb{R}^{1,3} \oplus \mathfrak{u}(1,3)$, which generalizes the Poincaré algebra $\mathbb{R}^{1,3} \oplus \mathfrak{o}(1,3)$. The same procedure applies to any element of $\mathbb{R}^{1,3} \oplus \mathfrak{u}(1,3)$.

\subsection{Colored metric}

As a result, if we denote by $\mathbf{\hat A}  := \mathbf{A} - \mathbf{A}(0)$ the relative gauge with coordinates $\hat A_{\mu} := \mathbf{\hat A} \bigcdot\gamma_\mu$, the final metric components are \begin{eqnarray}  \nonumber
g_{\mu\nu}  =  \hat \gamma_\mu  \bigcdot\hat \gamma_\nu & = & \eta_{\mu\nu}\mathbf{1}_4  - \kappa \hat A_{\mu} \bigcdot\hat A_{\nu} +  \kappa A_{\mu}(0)\bigcdot A_{\nu}(0) \\ \label{eq:exactsolution}
  & = & \eta_{\mu\nu} \mathbf{1}_4  - \kappa A_{\mu}\bigcdot A_{\nu} +  2q A_{(\mu}   u_{\nu)}/m,
\end{eqnarray} where $A_\mu = \mathbf{A} \bigcdot\gamma_\mu$ is a $\mathrm{U}(1, 3)$ gauge potential (boson), and the last term represents a \textit{gravitational source} at the potential origin. The first perturbation term, $A_{\mu}\bigcdot A_{\nu}$, corresponds to a \textit{gravitation spacetime} linked to a pair of entangled bosons (i.e., a candidate for a \textit{graviton}).

Given that $ A \in \mathfrak{u}(1, 3) $ and $ u \in \mathbb{R}^{1,3} $, \textcolor{black}{fluctuations} of the complexified metric can be \textcolor{black}{symbolically} expressed as
\begin{equation*}
g \sim \eta + {A} \otimes {A} \sim  \eta + (u \oplus A) \otimes (u \oplus A),
\end{equation*}
where $\eta$ is the background Minkowski metric.  These \textcolor{black}{fluctuations}, mapped through canonical tensor product isomorphisms, correspond to elements of the extended tensor space $\left( \mathbb{R}^{1,3} \oplus \mathfrak{u}(1, 3) \right) \otimes \left( \mathbb{R}^{1,3} \oplus \mathfrak{u}(1, 3) \right)$, significantly broadening the conventional scope of the tensor space $\mathbb{R}^{1,3} \otimes \mathbb{R}^{1,3}$.

\textbf{Geometrical description}. In this formulation, \textcolor{black}{there is an operational equivalence between the torsion 2-form (field strength of our extended TEGR) and the U$(1,3)$ field strength, which is geometrically interpreted as a curvature 2-form (Appendix \ref{ap:colored}, Eq.~\ref{eq:torson_curvature}). Here, the double-copy of gauge fields $A \in \mathfrak{u}$(1, 3) plays the role of coupled torsion fields. This allows us to metaphorically represent the extended metric} as a \textit{double helix structure}, formed by pairs $A \otimes A$ of entangled $\mathfrak{u}(1,3)$ vector fields\textcolor{black}{, instead of the classical rotated-transportation analogy of `spatial curves with torsion'}.

\textbf{Physical description}. The entangled\footnote{The term ``entangled bosons'' is used metaphorically to describe coupled gauge field components $A_{(\mu} A_{\nu)}$ in the \textit{linearized} gravity $g \sim\eta + A\otimes A$. While evocative of quantum language, the theory remains classical here, and the graviton is interpreted as an effective geometric excitation that requires future quantum interpretations in a separate paper} pairs are physically interpreted as \textit{bosons} (e.g., photons in QED), which act as virtual particles facilitating interactions within the extended symmetry group of spacetime. \textcolor{black}{In a more general sense}, these fields correspond to the connection of a double-copy gauge transformation governed by an extended Poincaré algebra $\mathbb{R}^{1,3} \oplus \mathfrak{u}(1,3)$.

\textbf{Classical limit}. When \textcolor{black}{metric fluctuations} are restricted to diagonal components, the resulting metric generates a $Cl_{4}(\mathbb{C})$ algebra defined by $\{\hat{\gamma}_\mu\}_{\mu = 0}^3$. This framework recovers classical geometrical structures, such as the Kaluza--Klein metric and the Kerr–Schild form (or their Kerr--Schild--Kundt perturbations), where $g \sim\eta + A \otimes A$. Prominent examples include the Kerr--Newman and Reissner--Nordström black holes \cite{Monteiro2021}. 

\textcolor{black}{Therefore, we need to distinguish between \textit{simple} and \textit{extended} gravity. For simple (or classical) gravity, only \textit{identity}-proportional terms in $A \otimes A$ can effectively interact with the classical sector of the metric $\mathbf{1}_4\eta_{\mu\nu}$ and velocity $\mathbf{1}_4u_{\mu}$ components by injection $\mathbf{1}_4 a \mapsto a \in \mathbb{R}$. In this sense, classical gravity is an Abelian case of deformations in the extended metric via the U(1,3) gauge symmetry, like electromagnetism \citep{Monjo2024}. The only difference is the double copy of the gauge field $A_{(\mu} A_{\nu)}$ required in the metric tensor (spin 2) versus the unique field $A_{\mu}$ for the limit approaching electromagnetism. Therefore, the graviton is not a fundamental particle in this proposed framework, but it is a bilinear composition of gauge bosons (e.g. photons), which should be further explored to see whether they are extended to non-Abelian sectors at unification energy scales.}

\textcolor{black}{At the current stage of the work presented in the following sections, all developments of SM embedding are limited to the first generation of fermions, excluding graviton interactions. However, extensions of the model could be explored with horizontal (flavor) symmetries from the second part of the double copy mentioned above, enabling a composed hypercolor-flavor graviton. Although interesting, these points are beyond the scope of this article.}

\section{Embedding of SM generators}
\label{sec:embedding_SM}

%\subsection{Generators of $\text{U}(1,3)$}

For the matrix representation of the $\mathfrak{u}$(1,3) algebra, we select an orthonormal base $B$ of four elements so called 1 `lepton' ($l$) and 3 `colors' ($r, g, b$) to identify with the (1,3) signature of the metric, which is $B = \{\ket{l}, \ket{r}, \ket{g}, \ket{b}\}$ and their corresponding dual basis $B^\dagger = \{\bra{\bar{l}}, \bra{\bar{r}}, \bra{\bar{g}}, \bra{\bar{b}}\}$. The group linked to this algebra is acting on the four-dimensional Dirac spinors  $\psi= (\psi_L,\psi_R)$, where $\psi_L$ and $\psi_R$ are two-component Weyl spinors. We then consider the \textit{state} configuration $\ket{\Psi}$ of a multiplet $\Psi$ consisting of four Dirac spinor fields $\Psi = (\psi_1, \psi_2, \psi_3, \psi_4)_B$, each with two possible states, \textit{up} and \textit{down} ($\ket{\Uparrow}, \ket{\Downarrow}$) for both the lepton and color elements, and $\ket{0}$ represents an empty state \footnote{To become familiar with this selection, one can consider the \textit{isospin} to interpret the \textit{up} and \textit{down} states of the lepton element as the \textit{neutrino} ($\nu$) and the \textit{electron} ($e$), respectively, while the states for the colors correspond to the \textit{quark flavors}, up ($u$) and down ($d$).}. For instance, consider the following initial up-down balanced configuration (with normalization omitted): \begin{equation}
   \ket{\Psi} =\begin{pmatrix}
\ket{\Downarrow} \\
\ket{\Uparrow} \\ 
\ket{\Downarrow}\\
\ket{\Uparrow} 
\end{pmatrix}_B = \underbrace{\ket{\Downarrow} \otimes \ket{l}}_{\Downarrow_l} +  \underbrace{\ket{\Uparrow} \otimes \ket{r}}_{\Uparrow_{r}} + \underbrace{\ket{\Downarrow} \otimes \ket{g}}_{\Downarrow_{g}} + \underbrace{\ket{\Uparrow} \otimes \ket{b}}_{\Uparrow_{b}}\,.
\end{equation}

Let $\{d\varphi_I(x)\}_{I=0}^{15}$ represent a set of 16 infinitesimal \textit{angles} or real phases of $\Psi$ at the position $x \in \mathbb{R}^{1,3}$. Then, consider it as a transformation produced by the unitary operator $U(x) = \exp({{{i}}d\varphi_I(x) \ell^a}) \in \text{U}(1,3)$, where $\{\ell^a\}_{I=0}^{15}$ is the set of generators of the non-compact $\text{U}(1,3)$ group. In other words, they allow its unitary relation $U^{\dagger}(x) \eta_{1,3} U(x) = \eta_{1,3}$ with Hermitian adjoint operator $U^{\dagger}(x) := \exp({-{{i}}d\varphi_I(x) \ell^a})$, and therefore satisfy the hermiticity condition $(\ell_I)^\dagger = \eta \ell_I \eta^{-1}$. Therefore, these generators are 15 traceless (i.e. ensuring $\det U = 1$) from $\text{SU}(1,3) \subset \text{U}(1,3)$ and 1 generator equivalent to the identity from $\text{U}(1)$ symmetry. 

%Using this basis, the eigenvectors depend on the spinor fields $\psi$ considered. 

 To display a matrix representation, we choose the scaling $\tfrac{1}{2}$ in order to obtain a trace-relation normalization of $\textbf{Tr}(\ell_I\ell_j)=\frac{1}{2}\delta_{IJ}$. They can be classified into five categories: \begin{enumerate}
 \item \textbf{$\text{U(1)}_{\Gamma}$ \textcolor{black}{simple} symmetry}: The identity generator $\ell_0 = \Gamma_0$ linked to a U(1) gauge potential $B^\mu \to A^\mu_{U(1)} = B^\mu \Gamma_0$ is
\begin{equation}
\label{eq:identity}
\begin{matrix}
        \Gamma_0  = \tfrac{1}{2\sqrt{2}} \begin{pmatrix}
1 & 0 & 0 & 0 \\
0 & 1 & 0 & 0 \\
0 & 0 & 1 & 0 \\
0 & 0 & 0 & 1
\end{pmatrix} 
%% \\
= \tfrac{1}{2\sqrt{2}}(\ket{l}\bra{\bar l} + \ket{r}\bra{\bar r} + \ket{g}\bra{\bar g} + \ket{b}\bra{\bar b}),
\end{matrix}
\end{equation} This generator is responsible of the classic gravity within the $\text{U}(1,3)$ colored gravity (Eq. \ref{eq:exactsolution}).

\item \textbf{$\text{SU(3)}_{\lambda}$ \textcolor{black}{strong} symmetry}:  A subset of 8 objects $\{\ell_a\}_{a=1}^8$ corresponds to the generators of the compact subgroup in $\text{SU}(1,3)$, which can be located in the lower $ 3 \times 3 $ block of the matrices that are \textbf{totally lepton-decoupled} and mimic the $\text{SU}(3)$ Gell-Mann matrices $\{\lambda_{a} \}_{a=1}^8$ and their gauge potentials $\{G^\mu_a\}_{a=1}^8$ to represent $A^\mu_{SU(3)} = \sum_{a=1}^8 G^\mu_a {\lambda}_a$. Abusing the notation, we will say that  $\{\ell_a\}_{a=1}^8 \equiv \{\lambda_{a} \}_{a=1}^8$ although the last three where conveniently modified:
\begin{equation} 
\label{eq:gluons}
\begin{matrix}
\lambda_1 = \tfrac{1}{2}\begin{pmatrix}
0 & 0 & 0 & 0 \\
0 & 0 & 1 & 0 \\
0 & 1 & 0 & 0 \\
0 & 0 & 0 & 0
\end{pmatrix}, \;
\quad
\lambda_2 = \tfrac{1}{2}\begin{pmatrix}
0 & 0 & 0 & 0 \\
0 & 0 & -i & 0 \\
0 & i & 0 & 0 \\
0 & 0 & 0 & 0
\end{pmatrix}, \; 
%\\
\quad \quad
\lambda_3 = \tfrac{1}{2}\begin{pmatrix}
0 & 0 & 0 & 0 \\
0 & 1 & 0 & 0 \\
0 & 0 & -1 & 0 \\
0 & 0 & 0 & 0
\end{pmatrix}, \;
\\ 
\lambda_4 = \tfrac{1}{2}\begin{pmatrix}
0 & 0 & 0 & 0 \\
0 & 0 & 0 & 1 \\
0 & 0 & 0 & 0 \\
0 & 1 & 0 & 0
\end{pmatrix},
\quad
\lambda_5 = \tfrac{1}{2}\begin{pmatrix}
0 & 0 & 0 & 0 \\
0 & 0 & 0 & -i \\
0 & 0 & 0 & 0 \\
0 & i & 0 & 0
\end{pmatrix},  \;
%\\ 
\hspace{10mm}
\lambda_6 = \tfrac{1}{2\sqrt{3}} \begin{pmatrix}
0 & 0 & 0 & 0 \\
0 & 1 & 0 & 0 \\
0 & 0 & 1 & 0 \\
0 & 0 & 0 & -2
\end{pmatrix}, \;
\\
\lambda_7 = \tfrac{1}{2}\begin{pmatrix}
0 & 0 & 0 & 0 \\
0 & 0 & 0 & 0 \\
0 & 0 & 0 & 1 \\
0 & 0 & 1 & 0
\end{pmatrix},\;
\quad
\lambda_8 = \tfrac{1}{2}\begin{pmatrix}
0 & 0 & 0 & 0 \\
0 & 0 & 0 & 0 \\
0 & 0 & 0 & -i \\
0 & 0 & i & 0
\end{pmatrix}. 
& {}
\end{matrix}
\end{equation}

\item \textbf{One $\text{U(1)}_{y}$ \textcolor{black}{hypercharge} symmetry}: A non-compact representative $y_w$ of $su(1,3)$ is also diagonal like $\lambda_6$ and $\lambda_3$, and it is isomorphic to the unit:
%https://en.wikipedia.org/wiki/W_and_Z_bosons#Predictions_of_the_W+,_W%E2%88%92_and_Z0_bosons
\begin{equation}
y_w = \tfrac{1}{2\sqrt{6}} \begin{pmatrix}
-3 & 0 & 0 & 0 \\
0 & 1 & 0 & 0 \\
0 & 0 & 1 & 0 \\
0 & 0 & 0 & 1 
\end{pmatrix}.
\end{equation} Like the $\Gamma_0$, the operator $y_w$ is coupled to the lepton sector (i.e. first row/column) \textcolor{black}{as a part of the electroweak interaction}. At this point we define total U(1) symmetry as $\left(\text{U}(1)_\Gamma \times \text{U}(1)_y\right) / \mathbb{Z}_2$, with the following $\mathfrak{u}(1)$ algebra representative:\begin{equation}
\label{eq:chi_u1}
    \chi^{\pm} = \pm\tfrac{1}{2}(\sqrt{2}\Gamma_0 + \sqrt{6}y_w) = \tfrac{1}{2} \begin{pmatrix}
-1 & 0 & 0 & 0 \\
0 & 1 & 0 & 0 \\
0 & 0 & 1 & 0 \\
0 & 0 & 0 & 1 
\end{pmatrix}.
\end{equation}This object recovers the signature of the spacetime and, at rest, induces a symmetry to keep the spatial components together. On the other hand, the $y_w$ operator is also fundamental for defining six completely-diagonal operators:\begin{eqnarray} \label{eq:z1}
  z_{\downarrow\uparrow\downarrow\uparrow}^{\pm} &:= & \pm \tfrac{1}{3}(\sqrt{6}y_w - \sqrt{3}\lambda_6 +3\lambda_3)
    \\  \label{eq:z2}
    z_{\downarrow\downarrow\uparrow\uparrow}^\pm &:=& \pm \tfrac{1}{3}(\sqrt{6}y_w - \sqrt{3}\lambda_6 -3\lambda_3)
    \\    \label{eq:z3}
    z_{\downarrow\uparrow\uparrow\downarrow}^\pm &:=&  \pm \tfrac{1}{3}(\sqrt{6}y_w + 2\sqrt{3}\lambda_6)
\end{eqnarray} \textcolor{black}{where $\text{Tr}(z_{\downarrow\uparrow\downarrow\uparrow}^\pm z_{\downarrow\uparrow\downarrow\uparrow}^\pm) = \text{Tr}( z_{\downarrow\downarrow\uparrow\uparrow}^\pm z_{\downarrow\downarrow\uparrow\uparrow}^\pm) = \text{Tr}(z_{\downarrow\uparrow\uparrow\downarrow}^\pm z_{\downarrow\uparrow\uparrow\downarrow}^\pm) = 1 = \text{Tr}(\chi^\pm\chi^\pm)$}. Due to its role in these equations, we define \textit{weak hypercharge} as \footnote{For left-chiral fermions $(e_L, u_L, d_L, u_L)$, the weak hypercharge will be $Y_W = \tfrac{1}{2}(-1, \tfrac{1}{3}, \tfrac{1}{3}, \tfrac{1}{3})$ in this paper.}\begin{equation}
\label{eq:hypercharge}
     Y_W := \tfrac{1}{3}\sqrt{6}y_w = \tfrac{1}{6} \begin{pmatrix}
-3 & 0 & 0 & 0 \\
0 & 1 & 0 & 0 \\
0 & 0 & 1 & 0 \\
0 & 0 & 0 & 1 
\end{pmatrix}\,.
\end{equation} 

Operators of Eq. \ref{eq:z1}--\ref{eq:z3} are useful to obtain eigenvalues of the multiplets. For instance, consider $\ket{\Psi} = \ket{\Psi^{+}} :=(\ket{\Downarrow}, \ket{\Uparrow}, \ket{\Downarrow}, \ket{\Uparrow})_B$, or $\ket{\Psi} = \ket{\Psi^{-}} :=(\ket{\Uparrow},\ket{\Downarrow},\ket{\Downarrow}, \ket{\Uparrow})_B$\footnote{Following the same familiar example above, $\ket{\Psi^{+}}= (\ket{\Downarrow}, \ket{\Uparrow}, \ket{\Downarrow},\ket{\Uparrow})_B$ corresponds to a proton ($\ket{\Uparrow}, \ket{\Downarrow}, \ket{\Uparrow}$) together with an electron ($\ket{\Downarrow}$), while \textcolor{black}{$\ket{\Psi^{-}}= (\ket{\Uparrow}, \ket{\Downarrow}, \ket{\Uparrow},\ket{\Downarrow})_B$ is a neutron ($\ket{\Downarrow}, \ket{\Uparrow},\ket{\Downarrow}$)} with a neutrino ($\ket{\Uparrow}$)}. For $\ket{\Psi^{+}}$, we will use $z_{\downarrow\uparrow\downarrow\uparrow}^+$ since its eigenvectors are $\Uparrow_l := (\ket{\Downarrow}, \ket{0},\ket{0},\ket{0})$, $\Uparrow_{r} := (\ket{0},\ket{\Uparrow},\ket{0},\ket{0})$, $\Downarrow_{g} := (\ket{0},\ket{0},\ket{\Downarrow},\ket{0})$, and $\Uparrow_{b} := (\ket{0},\ket{0},\ket{0}, \ket{\Uparrow})$, so
\begin{equation}\begin{matrix}
  z_{\downarrow\uparrow\downarrow\uparrow}^+  \Downarrow_l  &= -\tfrac{1}{2} \Downarrow_l, 
   \\
   z_{\downarrow\uparrow\downarrow\uparrow}^+  \Uparrow_{r}  &= +\tfrac{1}{2} \Uparrow_{r}, 
    \\
    z_{\downarrow\uparrow\downarrow\uparrow}^+ \Downarrow_{g}  &= -\tfrac{1}{2} \Downarrow_{g},
    \\
    z_{\downarrow\uparrow\downarrow\uparrow}^+ \Uparrow_{b}  &= +\tfrac{1}{2} \Uparrow_{b}.
\end{matrix}\end{equation} Therefore, for our initial configuration $\ket{\Psi^{+}}= (\ket{\Downarrow}, \ket{\Uparrow}, \ket{\Downarrow},\ket{\Uparrow})_B$, we identify the \textit{third weak isospin component} $T_3 = z_{\downarrow\uparrow\downarrow\uparrow}^+$ and the \textit{electric charge} can be finally defined as $Q = T_3 + Y_W$ \footnote{In our case, the factor $\tfrac{1}{2}$ of the SM equation $Q = T_3 + \tfrac{1}{2}Y_W$ is absorbed by the definition of $Y_W$ and the electric charge of $\ket{\Psi^{+}}$ is then $Q(\ket{\Psi^{+}}) = (-1, +2/3, -1/3, +2/3)$, as expected.}. %These definitions of \textit{quantum numbers} can be also extended to operators $X$ according to the change caused by it when applying to a vector $V$ (i.e. $\Downarrow_l,\Uparrow_r,\Downarrow_g$ or $\Uparrow_b$) of the initial configuration:\begin{eqnarray}
   % T_3(X) &:=& T_3(X\,V) - T_3(V) \\
   % Y_W(X) &:=& Y_W(X\,V) - Y_W(V) \\
   %Q(X) &:=& T_3(X) + Y_W(X)
%\end{eqnarray}

\item \textbf{First part of $3\times \text{SU(2)}_{w}$ \textcolor{black}{weak} symmetries}: Also interacting with the lepton sector, another subset of 3 non-compact rotation generators can be identified among the $\text{SU}(1,3)$ generators as follows, one per each color linked:
\begin{equation} 
\label{eq:w} 
\begin{matrix}
w_{1,r}^{\pm} = \pm\tfrac{1}{2}\begin{pmatrix}
0 & 1 & 0 & 0 \\
-1 & 0 & 0 & 0 \\
0 & 0 & 0 & 0 \\
0 & 0 & 0 & 0
\end{pmatrix},\;
w_{1,g}^{\pm} = \pm\tfrac{1}{2}\begin{pmatrix}
0 & 0 & 1 & 0 \\
0 & 0 & 0 & 0 \\
-1 & 0 & 0 & 0 \\
0 & 0 & 0 & 0
\end{pmatrix}, \;
w_{1,b}^\pm = \pm\tfrac{1}{2}\begin{pmatrix}
0 & 0 & 0 & 1 \\
0 & 0 & 0 & 0 \\
0 & 0 & 0 & 0 \\
-1 & 0 & 0 & 0
\end{pmatrix},
\end{matrix}
\end{equation}
that can be physically interpreted as responsible of SU(2) \textit{decay procedures}, since they have correspondence with three ladder operators: \begin{eqnarray}
        T_{r}^{\pm} &=& w_{1,r}^{\pm} \pm \tfrac{1}{2}(\sqrt{3}\lambda_6-\lambda_3),\\
        T_{g}^{\pm} &=& w_{1,g}^{\pm}\pm \tfrac{1}{2}(\sqrt{3}\lambda_6+\lambda_3), \\
        T_{b}^{\pm} &=& w_{1,b}^{\pm}\pm\lambda_3.
\end{eqnarray}The $T_{r}^{\pm}$ and $T_{b}^{\pm}$ operators can apply to initial-configuration eigenvalues $\pm(-\tfrac{1}{2}, +\tfrac{1}{2}, -\tfrac{1}{2}, +\tfrac{1}{2})$, while $T_{b}^{\pm}$ applies to $\pm(-\tfrac{1}{2}, -\tfrac{1}{2},  +\tfrac{1}{2}, +\tfrac{1}{2})$. For example, consider the operator $T_{r}^{\pm}$ acting on $\ket{\Psi^{\pm}}$, so\begin{equation*}
\begin{matrix}
    T_{r}^{+}\ket{\Psi^{+}} = T_{r}^{+}(\ket{\Downarrow}, \ket{\Uparrow}, \ket{\Downarrow},\ket{\Uparrow})_B \to  
    \\
    {} \quad\quad\quad\quad\quad\quad\quad\quad
    \to (\ket{\Uparrow}, \ket{\Downarrow}, \ket{\Downarrow},\ket{\Uparrow})_B = \lambda_7\ket{\Psi^{-}}
    \\
    {}
    \\
    T_{r}^{-}\ket{\Psi^{-}} = T_{r}^{-}(\ket{\Uparrow}, \ket{\Downarrow},  \ket{\Uparrow},\ket{\Downarrow})_B \to
     \\
    {} \quad\quad\quad\quad\quad\quad\quad\quad
    \to (\ket{\Downarrow}, \ket{\Uparrow}, \ket{\Uparrow},\ket{\Downarrow})_B= \lambda_7\ket{\Psi^{+}},
    \end{matrix}
\end{equation*}
where the sign $+$ or $-$ only depends on the initial configuration of $\ket{\Psi^{\pm}}$\footnote{Notice that our initial configuration does not allow the action of $T_{g}^{\pm}$ as the \textit{green} position is not \textit{up}}. The corresponding eigenvalues (action of $z_{\downarrow\uparrow\downarrow\uparrow}^\pm$) have been omitted by simplicity.
%where we can identify two raising operators $T_i^{+} := - iK_i + N_i$ and other two lowering operators $T_i^{-} := - iK_i - N_i$ for $i=1,3$; while Higgs fields would be related to $\tfrac{1}{2}(T^{+}+T^{-}) = -iK_i$. 

%\; \text{or} \;T_3' = \tfrac{1}{2\sqrt{6}} \begin{pmatrix} -3 & 0 & 0 & 0 \\ 0 & 1 & 0 & 0 \\ 0 & 0 & 1 & 0 \\ 0 & 0 & 0 & 1 \end{pmatrix}.

 \item \textbf{Second part of $3\times\text{SU(2)}_{w}$ \textcolor{black}{weak} symmetry}: Finally, a subset of 3 non-compact generators corresponds to boosts between lepton/time and color/spatial components:
\begin{equation} \label{eq:K_boosts}
\begin{matrix}
w_{2,r}^\pm = \pm \tfrac{1}{2}\begin{pmatrix}
0 & i & 0 & 0 \\
i & 0 & 0 & 0 \\
0 & 0 & 0 & 0 \\
0 & 0 & 0 & 0
\end{pmatrix}, \;
w_{2,g}^\pm = \pm \tfrac{1}{2}\begin{pmatrix}
0 & 0 & i & 0 \\
0 & 0 & 0 & 0 \\
i & 0 & 0 & 0 \\
0 & 0 & 0 & 0
\end{pmatrix}, \;
w_{2,b}^\pm = \pm \tfrac{1}{2}\begin{pmatrix}
0 & 0 & 0 & i \\
0 & 0 & 0 & 0 \\
0 & 0 & 0 & 0 \\
i & 0 & 0 & 0
\end{pmatrix}.
\end{matrix}
\end{equation}These generators are part of three subalgebras, $\mathfrak{su}(2)_r$, $\mathfrak{su}(2)_g$, and $\mathfrak{su}(2)_b$, isomorphic to $\mathfrak{su}(2)$, whose representative bases are $w_{c} = \{w_{0,c}^{\pm}, w_{1,c}^{\pm}, w_{2,c}^{\pm}\}$ for $c \in (r,g,b)$ and diagonal matrices $w_{0,c}^\pm$ defined from the hypercharge generator of the mixing $\text{U(1)}_{y}$ as follows:
\begin{eqnarray}
    w_{0,r}^\pm &=& z_{\downarrow\uparrow\downarrow\uparrow}^{\pm} \pm \tfrac{1}{2}(\sqrt{3}\lambda_6-\lambda_3) = \pm  \tfrac{1}{2}\begin{pmatrix}
    -1 & 0 & 0 & 0 \\
    0 & 1 & 0 & 0 \\
    0 & 0 & 0 & 0 \\  
    0 & 0 & 0 & 0
    \end{pmatrix},\;\;
    \\
    w_{0,g}^\pm &=& z_{\downarrow\downarrow\uparrow\uparrow}^\pm \pm \tfrac{1}{2}(\sqrt{3}\lambda_6+\lambda_3)= \pm  \tfrac{1}{2}\begin{pmatrix}
    -1 & 0 & 0 & 0 \\
    0 & 0 & 0 & 0 \\
    0 & 0 & 1 & 0 \\  
    0 & 0 & 0 & 0
    \end{pmatrix},\;\; 
    \\
    \label{eq:w0}
w_{0,b}^\pm &=&  z_{\downarrow\uparrow\downarrow\uparrow}^{\pm}\mp \lambda_3 = \pm  \tfrac{1}{2}\begin{pmatrix}
    -1 & 0 & 0 & 0 \\
    0 & 0 & 0 & 0 \\
    0 & 0 & 0 & 0 \\  
    0 & 0 & 0 & 1
    \end{pmatrix},\;\; 
\end{eqnarray} These three independent generators ($w_{0,r}^\pm$, $w_{0,g}^\pm$ and $w_{0,b}^\pm$) \textbf{complete the last part} of our triple $\mathfrak{su}(2)$ algebra: $\mathfrak{su}(2)_r,\, \mathfrak{su}(2)_g, \, \mathfrak{su}(2)_b$. However, the vector space of $\{w_{0,r}^\pm$, $w_{0,g}^\pm$, $w_{0,b}^\pm\}$ is equivalent to the generated by $\{\lambda_3, \lambda_6, y_w\}$, where two of them ($\lambda_3$ and $\lambda_6$) are already used in the $\text{SU(3)}_{\lambda}$ symmetry. Thus, only one generator $w_{0,c}^\pm$ (from the three colors $c \in \{r,g,b\}$) needs to be chosen to replace $y_w$ every time the $\text{SU(2)}_{w}$ symmetry acts.

Therefore, we need to reduce the triple $\mathfrak{su}(2)$ basis to just one effective algebra $\mathfrak{su}(2)_w^\text{eff}$ by removing the choice of color,\begin{equation}
   \mathfrak{su}(2)_w^\text{eff} := \text{choice}\{ \mathfrak{su}(2)_r,\, \mathfrak{su}(2)_g,\, \mathfrak{su}(2)_b\},
\end{equation} so we obtain the effective set of $\text{SU(2)}_{w}$ symmetry generators. Using any of the three equivalent bases $w_{c}$ with $c \in \{r,g,b\}$, the gauge potential can be defined as $A^\mu_{\text{SU}(2)} = W^\mu_a w_c^a$, where $w_c^a$ is an element of the basis $w_c = \{w_c^a\}_{a=1}^3$ for $c \in \{r,g,b\}$. If the colors are assumed to be equiprobable (e.g. in a lepton-lepton interaction), the effective basis of our $\mathfrak{su}(2)$ becomes a $\tfrac{1}{3}$-weighted mixture of \begin{equation}\label{eq:beta_mix}
\begin{matrix}
    \{ w_{2,r}^\pm , w_{2,g}^\pm, w_{2,b}^\pm;\,w_{1,r}^{\pm},w_{1,g}^{\pm},w_{1,b}^{\pm};\,z_{\downarrow\uparrow\downarrow\uparrow}^{\pm},\lambda_6^{\pm},\lambda_3^{\pm}\} =: w_l.
\end{matrix}
\end{equation}Notice that $z_{\downarrow\uparrow\downarrow\uparrow}^{\pm}$ itself, or originally represented by $y_w$, plays a central role in our $\mathfrak{su}(2)$ algebra by providing eigenvalues of the spinor multiplet, so it is expected to participate in all the cases. 
\end{enumerate}

Therefore, the mapping of the lepton-decoupled $\text{SU}(1,3)$ generators to the 8 representatives of the $\mathfrak{su}(3)$ algebra, plus the identification of the $\text{U}(1)_\Gamma$ generator and the effective $\text{SU}(2)_w$ illustrate the embeddings of $\text{SU}(3)$ and $\text{U}(1) \times \text{SU}(2)$ into $\text{U}(1,3)$. At this point, it is important to note that the purely color generators of $\text{SU}(3) \subset \text{SU}(1,3)$ do not directly interact with the lepton subspace\textcolor{black}{, so they do not require a dedicated symmetry-breaking mechanism}. Only during the procedure of the operators $\{w_{i,c}^{\pm}\}_{i=0}^2$ color generators play a role, which is just to balance the spatial region after disequilibrium caused by the rotation/boost between the lepton and color regions (which is interpreted as a \textit{decay procedure}). Due to the non-compact nature of the $\text{SU}(1,3)$ group, the lepton and hadron numbers are separately conserved \footnote{Therefore, free proton does not decay in this theory. Moreover, the electron-proton configuration ($\ket{\Psi^+}$) is expected to be more stable than the neutrino-neutron configuration ($\ket{\Psi^-}$) due to the electric charge dipole of the first.}. 

In contrast, the interactions between the diagonal generators $\Gamma_0$ and $z_{\downarrow\uparrow\downarrow\uparrow}^{\pm}$ introduce a potential mixing between the gauge potential $B_\mu$ of $\text{U}(1)_\Gamma$ and the $W^\mu_a$ bosons linked to $w_c^a$ in $\text{SU}(2)$. This results in weak interactions involving lepton-quark decays and supports the Higgs mechanism for mass generation (Sec.~\ref{sec:higgs_mechanism}). \textcolor{black}{Thus, the breaking $\text{U}(1,3) \to \text{SU}(3) \times \text{SU}(2) \times \text{U}(1)$ is also achieved through the Higgs mechanism as it preserves:  
\begin{itemize}[topsep=0pt,itemsep=0pt,parsep=0pt,partopsep=20pt]
\item The $\text{SU}(3)$ subgroup acting purely on colors (Eq.~\ref{eq:gluons}),  
\item The $\text{SU}(2) \times \text{U}(1)$ subgroup generated by $\{w_{i,c}^{\pm}, \chi^{\pm}\}$ (Eqs.~\ref{eq:w}--\ref{eq:w0} and Eq.~\ref{eq:chi_u1}),  
\end{itemize}
while breaking the remaining non-compact generators. This aligns with the SM gauge structure and does not necessitate extra symmetry-breaking assumptions beyond those implemented via the Higgs mechanism. In fact, although $\text{U}(1,3)$ has four more generators than the SM gauge group, these do not necessarily correspond to additional observable gauge bosons. The apparent redundancy is resolved through the color-linked embedding of $\text{SU}(2)$ interactions, where multiple copies of weak generators (e.g., $W^\pm$) act on different color directions. After spontaneous symmetry breaking, these degrees of freedom are effectively unified through an appropriate mixing (see Eq.~\ref{eq:beta_mix}), and the theory reproduces the SM boson spectrum without requiring additional physical gauge fields.}

 %Moreover, it is expected that the configuration $\ket{\Psi^+}$ is more stable than the configuration $\ket{\Psi^-}$ because of the electric charge dipole 
%, this embedding illustrates the mapping of $\text{SU}(1,3)$ generators to $\text{SU}(3)$ and $\text{U}(1) \times \text{SU}(2)$ components, demonstrating how the structure of the Standard Model arises within the $\text{U}(1,3)$ algebra. The lepton-decoupled generators directly connect to $\text{SU}(3)$, while weak mixing and diagonal operators provide the basis for $\text{U}(1) \times \text{SU}(2)$.

\section{The Higgs mechanism in $\text{U}(1,3)$}
% (responsible of the spontaneous breaking symmetry of the subgroup $\text{U}(2) \cong \text{SU}(2) \oplus \text{U}(1)$, as described in \textbf{Section XX})
\label{sec:higgs_mechanism}

\subsection{Embedding Higgs fields}

 Let $\phi_1(x)$, $\phi_2(x)$, $\phi_3(x)$ and $\phi_4(x)$ be four scalar fields involved in the Higgs mechanism. Under the $\text{U}(1,3)$ framework, it is expected that these scalar fields couple different components of the $\mathfrak{u}(1,3)$ multiplet and facilitate the spontaneous breaking of the $\text{SU}(2) \times \text{U}(1)$ symmetry down to a residual $\text{SU}(2)$ weak interaction and a $\text{U}(1)$ group identified with electromagnetism. %The interplay between group generators, scalar fields, and vacuum expectation values (VEVs) forms the basis of this framework.

Firstly, a scalar field matrix $\Phi(x)$ is constructed by excluding the rotation/boost operators ($w_{1,c},\,w_{2,c}$) and then by combining complementary generators of $\text{U}(1,3)$, which couple the lepton state $\ket{l}$ (associated with the time-like component) and the three color states $\{\ket{r}, \ket{g}, \ket{b}\}$ (associated with the spatial directions), multiplied by the four real scalar fields $\phi_1(x)$, $\phi_2(x)$, $\phi_3(x)$, and $\phi_4(x)$. The resulting scalar field matrix can be written in terms of four real matrices: two with diagonal elements, $M_1 := \sqrt{2}\Gamma_0 + \lambda_4$ and $M_2 := z_{\downarrow\uparrow\downarrow\uparrow}^{-} - \lambda_4$, and other two purely off-diagonal matrices, $M_3 := i\lambda_2 - i\lambda_8$ and $M_4 := \lambda_1 + \lambda_7$. Using these, we define four versions of $\Phi$, related to the interactions of $\Psi^{\pm} \leftrightarrow \Psi^{\pm}$ and $\Psi^{\pm} \leftrightarrow \Psi^{\mp}$:
\begin{equation} \label{eq:Phi}
\begin{matrix}
     \Phi_{++} := \phi_1\,M_1 + 
    \phi_2\,iM_2  +
    \phi_3\,M_1 + 
    \phi_4\,iM_2
    \\
         \Phi_{--} := \phi_1\,M_1 - 
    \phi_2\,iM_2  -
    \phi_3\,M_1 + 
    \phi_4\,iM_2
    \\
    \Phi_{+-} := \phi_1\,M_4 - 
    \phi_2\,iM_3 +
    \phi_3\,M_2 -
    \phi_4\,iM_1
    \\
        \Phi_{-+} := \phi_1\,M_4 - 
    \phi_2\,iM_3  -
    \phi_3\,M_2 + 
    \phi_4\,iM_1
\end{matrix}
\end{equation}Eq. \ref{eq:Phi} produces four $4 \times 4$ matrix representation of $\Phi(x)$ with nonzero entries in the  purely-color block and anothe nonzero element in the purely-lepton position, that is:

\begin{equation}\begin{matrix}
\Phi_{++}  = \frac{1}{2}
\begin{pmatrix}
\phi_0  & 0 & 0  & 0 \\
0 & \phi_0^*   & \phi_+ & \phi_0^*   \\
0  & \overline{\phi_+} & \phi_0  & \overline{\phi_+}  \\
0 & \phi_0^* & \phi_+ & \phi_0^*
\end{pmatrix},
\Phi_{--}  = \frac{1}{2}
\begin{pmatrix}
\phi_0^*  & 0 & 0  & 0 \\
0 & \phi_0 & \overline{\phi_+} & \phi_0 \\
0  & \phi_+ & \phi_0^*  & \phi_+ \\
0 & \phi_0 & \overline{\phi_+} & \phi_0
\end{pmatrix} 
\\
 \Phi_{+-}  = \frac{1}{2}
\begin{pmatrix}
\phi_+ & 0   & 0 & 0   \\
0  & \overline{\phi_+} & \phi_0  & \overline{\phi_+}  \\
0 & \phi_0^* & \phi_+ & \phi_0^* \\
0  & \overline{\phi_+} & \phi_0  & \overline{\phi_+} 
\end{pmatrix}  ,

\Phi_{-+}  = \frac{1}{2} \begin{pmatrix}
\overline{\phi_+} & 0   & 0 & 0   \\
0  & \phi_+ & \phi_0^{*}  & \phi_+  \\
0 & \phi_0^* & \overline{\phi_+} & \phi_0 \\
0  & \phi_+ & \phi_0^{*}  & \phi_+ 
\end{pmatrix}  
\end{matrix}\end{equation} where $\phi_+(x)$ and $\phi_0(x)$ are respectively the charged and the neutral components, conveniently defined as follows: \begin{equation}
\begin{matrix}
 \phi_0(x) = \phi_1(x) + i\phi_2(x), \quad \phi_+(x) = \phi_3(x) + i\phi_4(x) \\
  \phi_0^{*}(x) = \phi_1(x) - i\phi_2(x), \quad \overline{\phi_+}(x) = -\phi_3(x) + i\phi_4(x). \\
\end{matrix}
\end{equation}

Then, the components of $\Phi(x)$ can be mapped to a Higgs field doublet:
\begin{equation}
H(x) = \begin{pmatrix} \phi_+(x) \\ \phi_0(x) \end{pmatrix}
\end{equation} Here, the Higgs doublet transforms under a chosen $\text{SU}(2) \subset \text{U}(1,3)$ within the $\Phi$ structure (Eq. \ref{eq:Phi}), ensuring that $H(x)$ encodes the correct representation of the symmetry-breaking dynamics. The main role of the Higgs field is played by the scalar fields $\phi_1$ and $\phi_2$, which represent self-interactions in the spinor fields. These fields contribute to the symmetry-breaking process by introducing mass terms for gauge bosons. The fields $\phi_3$ and $\phi_4$ complete the definition of $\Phi(x)$ in Eq. \ref{eq:Phi}, coupling the time-like lepton state $\ket{l}$ and the space-like color states ${\ket{r}, \ket{g}, \ket{b}}$. These interactions are mediated by the non-compact generators of $\mathfrak{su}(1,3)$, providing a bridge between time-like and spatial elements.

All scalar fields involved in the Higgs mechanism could, in principle, be defined independently. However, their embedding into the symmetry algebra via the non-compact generators of $\mathfrak{su}(1,3)$ adds crucial value, allowing consistency within the group structure and defining their interactions with spinors and gauge bosons. This embedding enhances the theoretical framework by offering a consistent mathematical structure for symmetry breaking and interaction mediation, while also extending the standard Higgs mechanism.

%$P_L = \frac{1}{2}(1 - \gamma^5)$, $P_R = \frac{1}{2}(1 + \gamma^5)$

\subsection{Yukawa-like Lagrangian}

 In our framework, the spinoral states $\ket{\Psi^{-}} = \frac{1}{\sqrt{2}}(\Uparrow, \Downarrow, \Uparrow, \Downarrow)$ and $\ket{\Psi^{+}} =  \frac{1}{\sqrt{2}}(\Downarrow, \Uparrow, \Downarrow, \Uparrow)$\footnote{Now, with normalization factor $\frac{1}{\sqrt{2}}$ for convenient expression in the Lagrangian.} are organized similarly to the Dirac spinor multiplet $\Psi = (\psi_1, \psi_2, \psi_3, \psi_4)$. These spinoral states represent distinct configurations of particle flavors in the multiplet. If the left-hand and right-hand projectors are applied as $\Psi_L^{\pm} := P_L \Psi^{\pm}$ and $\Psi_R^{\pm} := P_R \Psi^{\pm}$, the interaction between $\Phi(x)$ and the spinors is given by terms of the form:
\begin{equation} \label{eq:yukawa1}
\mathcal{L}_{\text{Yukawa-like}} = -\bra{\bar{\Psi}_L^{\pm}} \Phi_{\pm \pm} y \ket{\Psi_R^{\pm}} + \bra{\bar{\Psi}_L^{\mp}} \Phi_{\pm\mp} y \ket{\Psi_R^{\pm}} + \text{h.c.}. %\tag{7}
\end{equation} Here, $\bar{\Psi}$ represents the Dirac adjoint of $\Psi$, ensuring proper Lorentz invariance of the interaction terms. The matrix $y = \text{diag}(y_1, y_2, y_3, y_4 )$ contains the Yukawa coupling constants, which control the strength of the interaction for each particle flavor. The Hermitian conjugate (h.c.) term ensures the Lagrangian is real, maintaining consistency with physical observables. This formulation highlights the interplay between the scalar field  and the spinoral multiplet, representing a fundamental mechanism for mass generation within the  framework.

For instance, the well-known first generation of the SM includes two flavors of leptons ($\nu$ and $e$) and two other flavors for quarks ($u$, $d$), corresponding to the states $\Uparrow$ and $\Downarrow$, respectively. Then, one can identify $\ket{\Psi^{+}} = \frac{1}{\sqrt{2}}(e, u, d, u)$ and $\ket{\Psi^{-}} = \frac{1}{\sqrt{2}}(\nu, d, u, d)$ and consequently Eq. \ref{eq:yukawa1} can be compactly summarized by using a unique pseudo\footnote{This is a pseudo-multiplet in the context of $\text{SU}(1,3)$ theory because is using a base with two leptons instead of just one like in the natural base of this theory.}-multiplet $\ket{{\hat \Psi}} = \frac{1}{\sqrt{2}}(\nu, e, u, d)$ as follows:
\begin{equation} \label{eq:pseudo_phi}
\begin{matrix}
\mathcal{L}_{\text{Yukawa-like}} =
 -\bra{\bar{\hat \Psi}_L} \hat \Phi(x) \hat{y} \ket{\hat \Psi_R}  + \text{h.c.},\;
 \\
 \quad\;\;
 \hat \Phi(x)  :=  \frac{1}{2}
\begin{pmatrix}
\phi_0^*  & \phi_+ & 0 & 0 \\
-\phi_+^* & \phi_0 & 0 & 0 \\
0 & 0 & \phi_0^*  & \phi_+ \\
0 & 0 & -\phi_+^* & \phi_0
\end{pmatrix}
\end{matrix}
\end{equation} where $\hat{y} = \text{diag}(y_\nu, y_e, y_u, y_d )$. Now, both the lepton and color regions are $2\times 2$ blocks. Moreover, it is assumed the possible existence of the sterile neutrino $\nu_R$ \cite{Barinov2022}.%, with a suppressing coupling factor $y_\nu$.
%(\nu_L\; e_L\; u_L\; d_L)
%\begin{pmatrix} \nu_R \\ e_R\\ u_R  \\ d_R  \end{pmatrix}

The Lagrangian describing the Higgs field and its interactions is given by $\mathcal{L}_{\text{Higgs}} = |D_\mu H|^2 - V(H)$, where the covariant derivative $D_\mu$ introduces the well-known gauge interactions $D_\mu H = \partial_\mu H - iq_{\text{SU}(2)} W_\mu H - iq_{\text{U}(1)} B_\mu H$ and the potential $V(H)$ can be constructed to allow spontaneous symmetry breaking, as usual, $V(H) = -\mu^2 H^\dagger H + \lambda (H^\dagger H)^2$. At the minimum of this potential, the Higgs doublet acquires a nonzero vacuum expectation value (VEV) such as $\langle H \rangle = \frac{1}{\sqrt{2}}( 0 , v )$, where $v = \sqrt{{-\mu^2}/{\lambda}}$ represents the energy scale at which the $\text{U}(1,3)$ symmetry breaks down to a residual $\text{U}(1)_\Gamma$. The charged component $\phi_+$ vanishes in vacuum, preserving charge conservation. 

This standard formulation, now with embedded scalars in $\Phi$, ensures that the Higgs mechanism provides masses to gauge bosons associated with broken symmetry generators while maintaining gauge invariance in the underlying framework. Additionally, the resulting scalar potential $V(H)$ naturally facilitates the emergence of a single physical scalar field, the Higgs boson, after symmetry breaking.

The spontaneous breaking of symmetry gives masses to the gauge bosons linked to the broken generators. Specifically, the ladder operators $T_{r}^\pm, T_{g}^\pm, T_{b}^\pm$, analogous to the $W^\pm$ bosons in the SM, acquire masses via their interactions with the Higgs VEV:
\begin{equation}
m_W = \frac{1}{2} q_{\text{SU}(2)} v.
\end{equation}
These operators mediate transitions between the lepton state $\ket{l}$ and the color states $\{\ket{r}, \ket{g}, \ket{b}\}$, forming the foundation for charged current interactions. Their mass terms arise from the interaction:
\begin{equation}
q_{\text{SU}(2)}^2 v h W^+_\mu W^{-\mu},
\end{equation}
where $h(x)$, the physical Higgs boson, represents fluctuations around the VEV:
\begin{equation}
H(x) = \frac{1}{\sqrt{2}} \begin{pmatrix} 0 \\ v + h(x) \end{pmatrix}.
\end{equation}
The Goldstone bosons corresponding to the broken symmetry generators are absorbed by the $W^\pm$ and $Z$ bosons \cite{Graf2017}, giving them mass while leaving one physical scalar degree of freedom, $h(x)$, in the spectrum. The $\Phi$ field is finally  
\begin{equation}\begin{matrix}
\Phi_{++}  = \Phi_{--} = \phi_1\cdot (\sqrt{2}\Gamma_0 + \lambda_4) 
\\
 \Phi_{+-}  = 
\Phi_{-+}  = \phi_1 \cdot (\lambda_1 + \lambda_7) 
\end{matrix}\end{equation}
where $\phi_1 = v + h(x)$ is the residual Higgs field, and $\{\lambda_a\}_{a\in\{1,4,7\}}$ encodes redundant information relative to $\sqrt{2}\Gamma_0$. This means that the information on spinor interactions encoded by $\Phi_{\pm\mp}$  is already embedded in the sub-blocks of $\Phi_{\pm\pm}$. Consequently, $\Phi_{\pm\pm} \sim \sqrt{2}\Gamma_0$ becomes the primary representative of the Higgs field in the $\text{U}(1,3)$ framework.  The Higgs boson, aside from color rotations, is determined by mixing the $B^\mu$ boson (linked to $\Gamma_0$).

\subsection{Prediction of the Weinberg angle}
\label{sec:predicted_angle}
\begin{figure*}
\centering
	\includegraphics[width=0.99\textwidth]{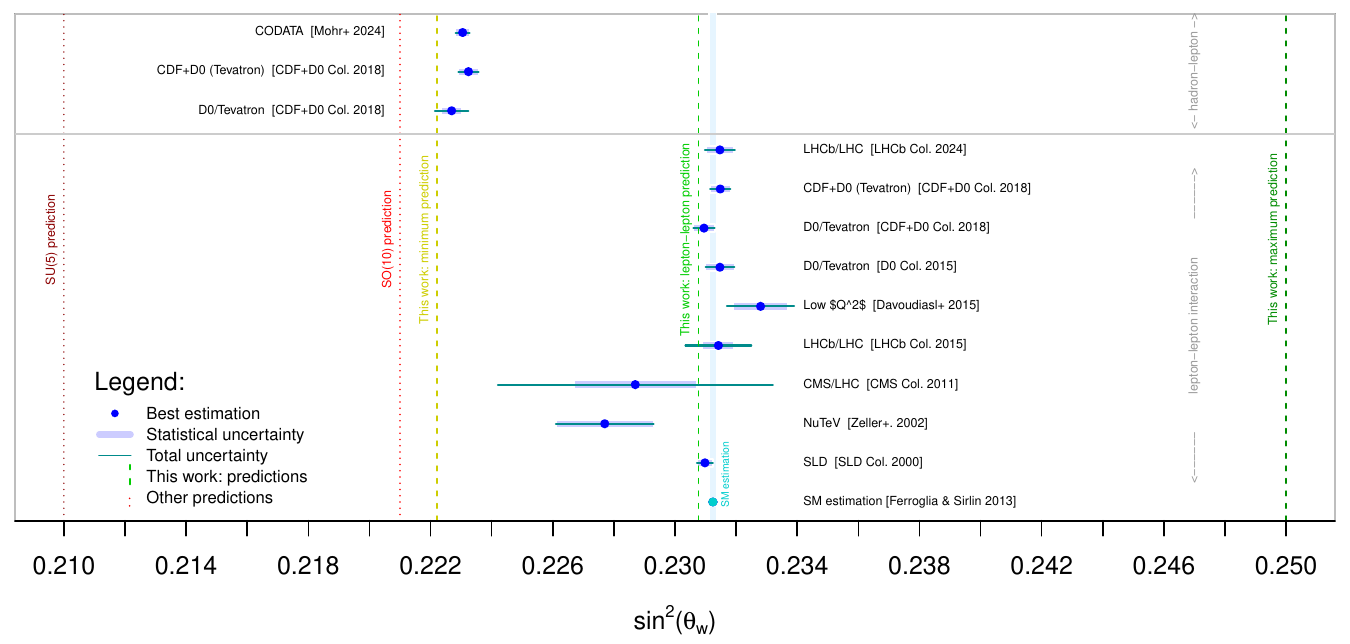}
    \caption{Comparison of theoretical predictions of $\sin^2\theta_W$ from colored gravity (dashed lines) and other GUT proposals (dotted lines), with respect empirical estimations (Table \ref{tab:Tab1}) represented in filled blue circles and uncertainty levels (horizontal lines). The SM prediction is added with its estimated uncertainty range (vertical light blue bland at $0.23152 \pm 0.00010$). The $\text{U}(1,3)$-based predictions are $0.2308$ (light green dashed line), which is separated on 0.5 to 2 $\sigma$ with respect to observations, and about $0.222$ (yellow dashed line) that is in $3\sigma$ tension. Maximum prediction corresponds to $\sin^2\theta_W = 0.25$ (dark green dashed line).
}  \label{fig:Fig01}
\end{figure*}

In this section, we provide details on a direct prediction of the weak mixing angle, or Weinberg angle, without considering quantum corrections. Theoretically, the Weinberg angle $\theta_W$ is related to the gauge coupling constants through: \begin{equation} \sin^2\theta_W = \frac{q_{\text{U}(1)}^2}{q_{\text{U}(1)}^2 + q_{\text{SU}(2)}^2}, \end{equation} where $q_{\text{U}(1)}$ and $q_{\text{SU}(2)}$ are the effective coupling constants for the U(1) and SU(2) gauge sectors, respectively, \textcolor{black}{derived from trace-based normalization ($f_{\text{U}(1)}$ and $f_{\text{SU}(2)}$) of their generators.}

Using the relationship between the coupling constants and their normalization, we factorize $q_{\text{U}(1)} = f_{\text{U}(1)} q_{\text{U}(1,3)}$ and $q_{\text{SU}(2)} = f_{\text{SU}(2)} q_{\text{U}(1,3)}$\textcolor{black}{, as two functions of the unknown \textit{coupling} constant $q_{\text{U}(1,3)}$ that describes the interaction strength with the spinor fields}. Consequently, the Weinberg angle can also be expressed in terms of $f_{\text{U}(1)}$ and $f_{\text{SU}(2)}$. It should be noted that the contributions from the configurations $+$ and $-$ are totally symmetric and produce exactly the same result. Considering only the $\text{SU}(2)$ basis $w_b$, the maximum expected value for the weak mixing angle is obtained: \begin{equation}\begin{matrix}
\label{eq:angle_W}
    \sin^2\theta_W^\text{max} = \frac{f_{\text{U}(1)}^2}{f_{\text{U}(1)}^2 + f_{\text{SU}(2)}^2} =  \frac{\text{Tr}(\Gamma_0\Gamma_0)}{\text{Tr}(\Gamma_0\Gamma_0) + \sum_{a=1}^3\text{Tr}(w_b^aw_b^a)} =
    \\
    = \frac{\tfrac{1}{2}}{\tfrac{1}{2} + \tfrac{3}{2}} =  \frac{1}{4}. 
\end{matrix} \end{equation}Thus, the predicted value of $\sin^2\theta_W = 0.25$ is reasonably close to the experimentally measured value of 0.231 (Table \ref{tab:Tab1}, Fig. \ref{fig:Fig01}). The discrepancy arises from the additional contributions from the $\text{U}(1,3)$ structure that might modify the coupling constants. \textcolor{black}{Specifically, the non-compact region provides three subgroups of SU(2)$\times$U(1) with a total of eight gauge fields that are mixed (Appendix~\ref{sec:adjoint_representation}): By internal symmetry, two of them are rotated, while six are merged in two effective bosons, that is, a third of the initial fields.}

\textcolor{black}{Therefore}, assuming complete mixing of the colors in lepton-lepton interactions, the effective basis of $\mathfrak{su}(2)$ is a $\tfrac{1}{3}$-weighted mixture of $w_l$ (Eq.~\ref{eq:beta_mix}) and the effective Weinberg angle is given by
\begin{equation} \begin{matrix}
    \sin^2\theta_W^\text{eff} =  \frac{\text{Tr}(\Gamma_0\Gamma_0)}{\text{Tr}(\Gamma_0\Gamma_0) + \tfrac{1}{3}\sum_{a=1}^9\text{Tr}(w_l^a w_l^a)} =
    \\
    = \frac{\tfrac{1}{2}}{\tfrac{1}{2} + \tfrac{1}{3}(8\cdot\frac{1}{2}+1)} =  \frac{6}{26} \approx 0.2308.
\end{matrix} \end{equation}  Here, all the basis elements of $w_l$ contribute $\tfrac{1}{2} = \text{Tr}(w_l^a w_l^a)$ except $z_{\downarrow\uparrow\downarrow\uparrow}^{\pm}$, which contributes $1$ (Eq.~\ref{eq:z1}). The resulting value of $0.2308$ is in slight tension (between 0.5 and 2 $\sigma$) with respect to the empirical estimations based on lepton-lepton experiments. 

On the other hand, the minimum expected value when two colors are mixed would be 
\begin{equation} \begin{matrix}
    \sin^2\theta_W^\text{min} 
    = \frac{\tfrac{1}{2}}{\tfrac{1}{2} + \tfrac{1}{2}(5\cdot\frac{1}{2}+1)} =  \frac{2}{9} \approx 0.2222.
\end{matrix} \end{equation} This result is more aligned to the experiments based on weak interactions of quark-lepton processes (Table \ref{tab:Tab1} lower), but is in tension between 3 and 4 $\sigma$ with the best empirical estimations.

Consequently, exploring the role of quantum corrections is necessary to explore whether quantum corrections could reduce the theoretical tension between colored gravity and the experimental value of the Weinberg angle. \textcolor{black}{Moreover, the model also predicts that effective bosons ($W_1$ and $W_2$) could be split into a total of six charged $W$ bosons at high energies, with the only difference testable in the color direction. The masses of these $W$ bosons should be identical, so just the mediation (or not) of detectable gluons could distinguish their color direction before their final decay process.}

\subsection{\textcolor{black}{Impact on the Lorentz invariance.}} \label{sec:impact_lorentz}

\textcolor{black}{The deformation of the spacetime generators $\gamma_\mu$ via the gauge potential $A_\mu$ introduces an effective local frame $\hat{\gamma}_\mu$, which in turn \textcolor{black}{extends} the four-velocity $\hat{u}_\mu$. This might suggest a deviation from traditional Lorentz symmetry. However, the theory retains local Lorentz invariance in the generalized sense: the \textcolor{black}{extended} frames transform covariantly under local $\mathrm{U}(1,3)$ transformations. The spontaneous breaking induced by the Higgs mechanism affects internal gauge symmetries — such as SU(2) and SU(3) — but not the spacetime symmetry group itself. Therefore, the appearance of massive gauge bosons (W, Z, gluons) reflects internal symmetry breaking, not a violation of Lorentz invariance. Nonetheless, if components of $A_\mu$ were to develop a spacetime-dependent or directional structure, this could lead to spontaneous breaking of Lorentz symmetry with a potential quantum noncommutative interpretation. Investigating this possibility requires analyzing the dynamical behavior of $A_\mu$ beyond the classical level, a task we defer to future work.}

%On the other hand, the contribution of $\lambda_3$ and $\lambda_6$ is very small compared to the weight of the $\text{SU}(3)$ generators. Assuming that $\lambda_3$ and $\lambda_6$ are two elements of the nine-component $\text{SU}(2)$ mix basis ($w_l$) and normalizing with the trace of $\text{SU}(3)$, we obtain $\sin^2\theta_{SU(3)} \approx 0.027$.

%their embedding into the symmetry algebra via the non-compact generators of $\mathfrak{u}(1,3)$ adds crucial value, ensuring consistency with the group structure and defining their interactions with spinors and gauge bosons.
%are introduced in the theory as spacetime-dependent functions that

\section{Final remarks}
\label{sec:conclusions}

The complexification of the Minkowskian metric for spinor fields leads to a natural $\text{U}(1, 3)$ symmetry that perturbs the spacetime generators. The gauge potential for the $\text{U}(1,3)$ group can be decomposed as $A^\mu = A^\mu_a \ell^a$, where $\ell^a$ are the 16 generators of $\text{U}(1,3)$, and $A^\mu_a$ are the corresponding gauge fields. The embedding of \textcolor{black}{standard} $\text{SU}(3)$ into $\text{U}(1,3)$ is achieved by placing their generators in the spatial $3 \times 3$ blocks of the $4 \times 4$ matrix structure. On the other hand, the embedding of $\text{U}(1) \times \text{SU}(2)$ is not direct and requires interactions between specific subgroups and symmetry generators.

From the resulting hierarchy, the purely color generators of $\text{SU}(3) \subset \text{SU}(1,3)$ do not directly interact with the lepton subspace. Only during the procedure of the $\text{SU}(2)$ operators $\{w_{i,c}^{\pm}\}_{i=0}^2$, the color generators play a role of spatial rotations by balancing the disequilibrium caused by the lepton-color interactions.

The Higgs mechanism in $\text{SU}(1,3)$ extends the SM by embedding its symmetry-breaking structure into the larger gauge group. In this extended framework, the four scalar fields $\phi_1, \phi_2, \phi_3, \phi_4$ are organized into a $4 \times 4$ representation of the Higgs doublet that facilitates the breaking of symmetry of the electroweak interaction. The non-compact generators mediate the couplings necessary for this breaking, with the neutral component acquiring a VEV to drive the process. This approach retains the essential features of the standard Higgs mechanism while integrating it within a broader theoretical structure.

The predicted weak mixing angle $\sin^2\theta_W$ is aligned to two different cases: a purely lepton-lepton interaction (yielding a value statistically compatible with $0.231$), and a hadron-lepton interaction, which shows a tension of about 3$\sigma$ with the best experimental observations. Exploring quantum corrections could further refine these predictions and potentially resolve the discrepancies.

This paper represents the initial development of a novel GUT proposal, with a natural (spacetime-related) Lie algebra capable of integrating gravity and other fundamental forces. The new theoretical framework could solve some existing open issues. For example, unlike the SU(5) model, proton stability is naturally expected in the $\text{SU}(1,3)$ framework due to the lepton-exclusion properties of its compact SU(3) subgroup. \textcolor{black}{Moreover, while this work is limited to the first generation of fermions, the model could be extended to three generations via horizontal (flavor) rotations by taking advantage of the second part of the double-copy gauge group.}

Future research directions include exploring quantum corrections to the Weinberg angle, better understanding flavor family mixing, and further embedding the Standard Model's interactions into this extended framework. Additionally, the quantization of the spacetime metric, a critical challenge for any GUT when including gravity, will be evaluated in light of our recent theoretical advancements. These explorations will also investigate connections to quantum perspectives, such as the dynamics of causal structures and the implications of quantum nonlocality \cite{CastroRuiz2018}. The quantization of spacetime itself and its integration into a comprehensive theory remain a pivotal challenge for this approach, paving the way for a deeper understanding of the nature's fundamental forces. \textcolor{black}{Although the coordinates are spinor-valued in this formalism, they encode geometric structures via Clifford algebra while retaining a consistent classical limit. This direction offers a novel route to incorporate gravity and gauge forces into a single algebraic framework. Thus, future developments may clarify how standard model interactions and spacetime locality emerge from deeper spin-geometric principles.}

% %Developing a full quantum field theory based on such coordinates is an open challenge that will be pursued in future work.

%Future work will focus on better understanding the embedding of SM interactions in the proposed framework, especially in quantum corrections of the Weinberg angle and in the mixing of the different flavor families. Moreover, we will evaluate the viability of \textit{quantum gravity} that comprises all our developments carried out to date and that incorporates the Bell bound and other quantum perspectives such as the dynamics of causal structures \cite{CastroRuiz2018}. The quantization of spacetime and the metric itself is a big challenge for any GUT proposal that aims to include gravity.

%\subsubsection{Wide equations}
%The equation that 
%\begin{widetext}
%\begin{equation}
%xxx
%\end{equation}
%\end{widetext}

\vspace{1mm}

\begin{acknowledgments}
The author acknowledges the invaluable support of R.O. Campoamor-Stursberg and A. Rodríguez-Abella for reviewing the algebraic structures used in this paper. Their suggestions and encouragements have made this work possible. This paper was written during the author's affiliations with the Complutense University of Madrid, the Saint Louis University, and, more recently, the University of Alcalá.
\end{acknowledgments}

\appendix
%\section{Some title}
%Please always give a title also for appendices.

%\paragraph{Note added.} This is also a good position for notes added after the paper has been written. \texorpdfstring{U(1,3)}{U(1,3)}

\color{black}

\section{\textcolor{black}{Basics of U(1, 3) colored gravity}}
\label{sec:basics}

\subsection{U(1, 3) gauge group}
\label{sec:u13}

The group $\mathrm{U}(1,3)$ is a non-compact unitary group that preserves a Hermitian form with respect to the metric $\eta = \mathrm{diag}(1, -1, -1, -1)$ of signature $(1,3)$, and it is given by 16 generators $\{\ell_a\}_{a=0}^{15}$ (Sec.~\ref{sec:embedding_SM}) that span the Lie algebra $\mathfrak{u}(1,3)$. Explicitly, for an element $U(x) = \exp(i\varphi^a(x) \ell_a) \in \mathrm{U}(1,3)$ with real parameters $\varphi^a(x)$, the defining condition $U^\dagger \eta U = \eta$ is satisfied. In turn, this leads to the hermiticity condition for the generators, $(\ell_a)^\dagger = \eta \ell_a \eta^{-1}$.

%\widehat
The measurement invariance for spinor fields when transformed as $\Psi(x)\to \Psi(x)'=U(x)\Psi(x)$ (Sec.~\ref{sec:perturbedspinor}) leads to a covariant derivative $\nabla_{\mu}$ satisfying the (gauge) transformation $\nabla_{\mu}\Psi(x) \mapsto {(\nabla_{\mu}\Psi)'}(x) = U(x) \nabla_{\mu}\Psi(x)$. This gauge covariant derivative must be $\nabla_{\mu}= \mathbf{1}_4 \partial_{\mu} - i  q_{\text{U}(1,3)} A_{\mu}$ for some real coupling constant $q_{\text{U}(1,3)}$ that describes the interaction strength between the spinors $\Psi(x)$ and the gauge fields $A_\mu = A_\mu^a \ell_a$, which transfom as
\begin{equation*}
A_\mu \mapsto A'_\mu = U A_\mu U^{-1} + i q_{\text{U}(1,3)}^{-1}(\partial_\mu U) U^{-1}.
\end{equation*} For infinitesimal parameters $\varphi^a(x) \ll 1$, the group element approaches $U(x) \approx \mathbf{1}_4 + i\varphi^a(x)\ell_a$ and the (infinitesimal) adjoint transformation of the gauge field is:\begin{equation}
     A_\mu \mapsto A'_\mu = A_\mu + \delta A_\mu^a\ell_a + \mathcal{O}(\varphi^2)
     \quad\text{with}\;
     \delta A_\mu^a :=  q_{\text{U}(1,3)}^{-1} \partial_\mu \varphi^a + f^{abc} \varphi^b A_\mu^c,
\end{equation} where $f^{abc}$ are the U$(1,3)$ structure constants of the group that satisfy the antisymmetric relation $[\ell^a,\ell^b]=if^{abc}\ell^c$. The field strength tensor takes the usual YM form:
\[
\mathcal{F}_{\mu \nu} := i q_{\text{U}(1,3)}^{-1}[\nabla_\mu,\nabla_\nu] = \partial_\mu A_\nu - \partial_\nu A_\mu +  q_{\text{U}(1,3)} [A_\mu, A_\nu],
\]
While this is structurally analogous to gauge theories based on compact groups, care must be taken due to the non-compactness of $\mathrm{U}(1,3)$.

In particular, the Lie algebra of $\mathrm{U}(1,3)$ possesses an indefinite Killing form, defined by $\kappa_{ab} := \mathrm{Tr}\big(\mathrm{ad}(\ell_a) \circ \mathrm{ad}(\ell_b)\big)$, where $\mathrm{ad}(\ell_a) = [\ell_a,\, \cdot\, ]$ is the adjoint action. Consequently, inner products and traces in the Lagrangian must be handled with this indefinite structure in mind. Specifically, contractions involving the field strength components must respect the group metric $\mathcal{F}_{\mu \nu}^a \mathcal{F}^{a \mu \nu}  \longrightarrow \; \kappa_{ab} \mathcal{F}_{\mu \nu}^a \mathcal{F}^{b \mu \nu}$.

%TEGR Lagrangian (cf. Eq.~\eqref{eq:lagrangian_TEGR})
Thus, the YM Lagrangian density associated with the $\mathrm{U}(1,3)$ gauge theory, analogous in structure to the compact YM case, is given by
\begin{equation}
\label{eq:ym_lagrangian}
\mathcal{L}_{\text{YM}} = -\frac{1}{4} \kappa_{ab} \mathcal{F}_{\mu \nu}^a \mathcal{F}^{b \mu \nu} \ :=\  -\frac{1}{4} \mathcal{F}_{\mu \nu}^a \mathcal{F}^{\mu \nu}_a,
\end{equation}
where the components of the field strength tensor in the adjoint representation are
\[
\mathcal{F}_{\mu \nu}^a = \partial_\mu A_\nu^a - \partial_\nu A_\mu^a + q_{\text{U}(1,3)} f^{abc} A_\mu^b A_\nu^c.
\]

\subsection{Telleparallel gravity equivalent of GR (TEGR)}
\label{sec:teleparalell}
Teleparallel gravities constitute a class of theories in which gravitation is described not by curvature but by torsion\footnote{The torsion belongs to the connection choice (e.g. Weitzenböck connection $\Gamma$), not to the metric itself. The Levi‑Civita connection $\hat{\Gamma}$ for the same metric will still be torsion‑free ($T(\hat{\Gamma}) = 0$) but curved (nonzero Ricci scalar, $R(\hat{\Gamma}) \neq 0$, unless the space‑time is exactly flat). The equivalence between both approaches is given by $R(\hat{\Gamma}) = - T(\Gamma) + 2{\hat{\nabla}_\mu}T^\mu(\Gamma)$, where $\hat{\nabla}_\mu$ is the covariant derivative of the Levi‑Civita connection $\hat{\Gamma}$.}. An interesting case is the TEGR based on the Weitzenböck connection $\Gamma$. In this framework, the Riemann curvature tensor vanishes ($R^{\sigma}_{\ \rho \mu\nu}(\Gamma) = 0$), and the spin connection is assumed to be null ($\omega_\mu^{\ ab} = 0$). Despite this reformulation, the Euler--Lagrange equations of TEGR follow from an action that is dynamically equivalent to the Einstein--Hilbert action, differing only by a total divergence term \citep{Andrade2000, Krssak2019}.

Inspired by the structure of gauge theories, some authors have employed translational invariance to formulate TEGR in a gauge-like language. Nonetheless, this so-called \textit{translation-gauge formalism} does not meet the standard criteria of a full gauge theory, such as the presence of a principal bundle structure \cite{Fontanini2019}. Regardless of the mathematical limitations, we can adopt this gauge-like approach due to its appealing analogies with classical gauge theories in particle physics.

Let us consider a Lorentzian manifold $(M, \mathbf{g})$, and associate to each point in $M$ a copy of Minkowski space $\mathcal{M}$. The coordinates in this internal anholonomic space are denoted by $\{x^a\}_a$, giving rise to the \textit{Minkowski bundle} $\mathcal{M} \to M$ with local coordinates $\{x^\mu, x^a\}_{\mu,a}$.

The local action of the translation group on the Minkowski bundle is given by $x^a \mapsto \hat{x}^a = x^a + \epsilon^a(x^\mu)$, where the Lie algebra is generated by $\{\partial_a\}_a$, the derivatives with respect to the anholonomic coordinates. The dynamical fields of teleparallel gravity are 1-forms on $M$ valued in this Lie algebra—referred to as \textit{translational gauge-like potentials}—denoted $\phi = \phi_{\mu} dx^\mu$, with $\phi_\mu = \phi_{\mu}^{\ a} \partial_a$. Under an infinitesimal translation $\delta x^a = \epsilon^a(x^\mu)$, these potentials transform as $\hat{\phi}_{\mu}^{\ a} = \phi_{\mu}^{\ a} - \partial_\mu \epsilon^a$, that is, the action of the translation group extends naturally to this family of potentials. Moreover, the potential $\phi_\mu = \phi_\mu^{\ a} \partial_a$ induces a transformation on the vierbein (tetrad) components as follows (see e.g., \cite{Aldrovandi2006}):
\begin{equation} \label{eq:tetrad}
e_{\mu}^{\ a} = \partial_{\mu} x^{a} \quad \mapsto \quad \hat{e}_{\mu}^{\ a} = e_{\mu}^{\ a} + \phi_{\mu}^{\ a} = \partial_{\mu} x^{a} + \phi_{\mu}^{\ a}.
\end{equation}

This construction leads naturally to a fluctuated metric $g_{\mu\nu} :=  \hat e_\mu\bigcdot\hat e_\nu$, as well as to the definition of the gauge-like derivative and differential in TEGR:
\begin{eqnarray*}
   \partial_\mu  &\mapsto& D_\mu  = \hat{e}_{\mu} = \hat{e}_{\mu}^{\ a} \partial_a = \partial_\mu + \phi_\mu^{\ a} \partial_a = \partial_\mu + \phi_\mu,
   \\
dx^\mu  &\mapsto& d\hat{x}^\mu = dx^\mu - \phi^\mu_{\ a} dx^a.
\end{eqnarray*} The associated field strength is $F_{\mu\nu} = F_{\ \mu\nu}^{a} \partial_a$, where $F^a_{\ \mu\nu} = \partial_\mu \phi^a_{\ \nu} - \partial_\nu \phi^a_{\ \mu}$.

The TEGR Lagrangian density can be expressed either in terms of the torsion tensor or the field strength as follows \cite{Krssak2019,deAndradePereira1997}:
\begin{equation} \label{eq:lagrangian_TEGR}
\mathcal{L} = \frac{\hat{e}}{2\kappa} \left( \frac{1}{4}T^\rho_{\ \mu \nu} T_{\rho}^{\ \mu\nu} + \frac{1}{2}T^\rho_{\ \mu\nu}T^{\nu\mu}_{\ \ \rho} - T^\rho_{\ \mu\rho}T^{\nu\mu}_{\ \ \nu} \right)
= \frac{\hat{e}}{2\kappa} \left( \frac{1}{4}F^a_{\ \mu\nu} F_{a}^{\ \mu \nu} \right),
\end{equation}
where $\kappa = 8\pi G$, and the torsion tensor $T^\rho_{\ \mu\nu} = \Gamma^\rho_{\ \nu\mu} - \Gamma^\rho_{\ \mu\nu}$ is defined via the (flat) Weitzenböck connection $\Gamma^\rho_{\ \mu\nu} = \hat{e}^{\rho}_{\ a} \partial_\nu \hat{e}^{a}_{\ \mu}$, with $\hat{e} := \det(\hat{e}_\mu^{\ a}) = \sqrt{-g}$. This Lagrangian yields the same dynamical content as the Einstein--Hilbert Lagrangian $\mathcal{L}_{GR} = \frac{1}{2\kappa} \hat{e} R$, where $R$ is the Ricci scalar. The two Lagrangians differ only by a total divergence:
\[
\mathcal{L} - \mathcal{L}_{GR} = - \frac{1}{\kappa} \partial_\mu (\hat{e} \, T^{\nu\mu}_{\ \ \ \nu}).
\]
Finally, notice that the TEGR Lagrangian (Eq.~\ref{eq:lagrangian_TEGR}) is totally analog to the YM Lagrangian (Eq.~\ref{eq:ym_lagrangian}). Furthermore, that similarity is not casual as it is discussed in Appendix \ref{ap:colored}.

\subsection{From TEGR to U$(1,3)$ colored gravity}
\label{ap:colored}

This section summarizes the fundamentals of U$(1,3)$ colored gravity, which was originally developed in \citep{Monjo2024} as an extension of TEGR. Let $\Psi(x) =  \exp(-{{i}}\, m \gamma_\nu dx^\nu) \Psi_0$ be an evolving multiplet of Dirac spinors $\Psi_0$ at $x = (x^\mu) \in \mathbb{R}^{1,3}$ with mass $m>0$. The components $m_\mu = \mathbf{m} \bigcdot \gamma_\mu$ of its four-momentum $\mathbf{m}$ are given by the classical momenta $m_\mu = m u_\mu \mathbf{1}_4 := m \frac{dx_\mu}{d\tau} \mathbf{1}_4$ for a proper time $d\tau^2 := \eta_{\mu\nu} dx^\mu dx^\nu$ (see Sec.~\ref{sec:prel}).

The following definition of \textit{operational equivalences} ``$\simeq$'' is crucial to bridge between classical magnitudes and operators applied to the spinor multiplet $\Psi$(x):\begin{equation}
\label{eq:bridge3}
      a \simeq b \;\;\;\; \Longleftrightarrow  \;\;\;\;  \braket{a}_\Psi = \braket{b}_\Psi\;, 
\end{equation} where $\braket{c}_\Psi$ is the expected value of the operator $c$\footnote{Formally, the operator $c$ is assumed to be square-integrable in $L^2$, and its expectation value is defined as the real part of the inner product, $\langle c \rangle_\Psi := \mathrm{Re} \langle \bar\psi |\, c \,| \psi \rangle$, where $\bar\psi = \psi^\dagger \gamma^0$ is the Dirac adjoint in this Hilbert space. For example, classical quantum mechanics defines the density current as $u^\mu := \mathrm{Re} \bra{\bar\psi} \hat{U}^\mu \ket{\psi}$ with the momentum operator $\hat{U}^\mu = \frac{{i}}{m} \partial^{\mu}$ evaluated for some mass $m \in \mathbb{R}_{>0}$.}. For instance, this allows the Dirac equation to be written compactly as ${i} \gamma^\mu \partial_\mu \simeq m$, for $\Psi$ with mass $m$. 

Also enabled by Eq.~\ref{eq:bridge3}, the colored gravity is based on an \textit{extended teleparallel gauge} (ETG) displacement in the anholonomic coordinates, which is defined as
\begin{equation}
\label{eq:displacement}
    \delta x^a \approx - \left(\, \mathbf{1}_4\, \phi^{\ a}_{\nu}{}_{\text{scalar}}  + \phi^{\ a}_{\nu} \,\right)\delta x^\nu,
\end{equation}
with $\phi^{\ a}_{\nu}{}_{\text{scalar}}$ corresponding to the usual TEGR fluctuation, and $\phi^{\ a}_{\nu}$ denotes an extended gauge translation with respect to the one introduced in the Sec.~\ref{sec:teleparalell}. 

Then, consider the displacement $\delta x \in \mathbb{R}^{1,3} \oplus \mathfrak{u}(1,3)$ with $\phi^{\ a}_{\nu}{}_{\text{scalar}} \approx 0$ and let
\[
\delta x^a \approx - \phi^{\ a}_{\nu} \delta x^\nu =  - A^a \bigcdot \frac{q}{\mathbf{m}} \bigcdot \gamma_\nu\, \delta x^\nu \simeq {i}q\, A^{\ a}_{\nu} \delta x^\nu,
\]
be a $\mathrm{U}(1,3)$ gauge transformation, also interpreted as a non-Abelian phase $\theta(x) = \varphi^I(x)\ell_I = A(x) = A_\mu(x) dx^\mu$ that modifies spinor trajectories and hence promotes the (oscillatory) local coordinates $\{x^\mu\}$ to matrix-valued quantities. From the transformed tetrad fields (Eq~\ref{eq:tetrad}) the resulting metric is $g(x) \sim \eta - \kappa\, A(x)\bigcdot A(x)$, where its fluctuation is an extension in U$(1,3)\times$U$(1, 3)$.

With the ETG prescription, the Cartan covariant derivative coincides with the teleparallel gauge derivative $D_\mu$, which is operationally equivalent to the U$(1,3)$ gauge covariant derivative $\nabla_\mu$:
\begin{equation} \label{eq:covariant3}
D_\mu  
\approx e_\mu + \frac{q}{m} A_\mu u^a \partial_a 
\simeq \partial_\mu - {{i}}q A_\mu = \nabla_\mu.
\end{equation} This \textit{expectation equivalence}, $D_\mu \simeq \nabla_\mu$, holds for each \textit{particle-field} involved (i.e., in the sense of expectation values) and plays a central role in ensuring the consistency of the theory when non-Abelian gauge fields are introduced.

Given the operational equivalence of covariant derivatives, the torsion 2-form provides the field strength ($F^a_{\ \mu\nu} = \partial_\mu \phi^a_{\ \nu} - \partial_\nu \phi^a_{\ \mu}$) of the teleparallel translational gauge, behaving analogously to a curvature 2-form. Applying expectation values (Eq.~\ref{eq:bridge3}), we obtain the SU$(1,3)$ field strength -—geometrically interpreted as the curvature 2-form:
\begin{equation}
\label{eq:torson_curvature}
F_{\mu\nu} := F_{\ \mu\nu}^{a} \partial_a 
= [D_\mu, D_\nu] = (D_\mu \hat{e}_\nu^{\ a} - D_\nu \hat{e}_\mu^{\ a}) \partial_a 
\approx \frac{q}{m} \mathcal{F}_{\mu\nu} u^a \partial_a 
\simeq -{{i}}q\, \mathcal{F}_{\mu\nu},
\end{equation}
where $\mathcal{F}_{\mu\nu} := {{i}}q^{-1}[\nabla_\mu, \nabla_\nu]$. Furthermore, the torsion scalar gives rise to the corresponding YM scalar. Specifically, the torsion field components relate to the gauge field strength components via:
\begin{equation}
F^a_{\ \mu\nu} \simeq \frac{q}{m} \mathcal{F}_{\mu\nu} u^a \quad \mapsto \quad 
F^a_{\ \mu\nu} F_a^{\ \mu\nu} \simeq \frac{q^2}{m^2} \mathcal{F}_{\mu\nu} \mathcal{F}^{\mu\nu},
\end{equation}
which holds true for the spinor multiplet $\Psi$, leading to an operational equivalence between the TEGR Lagrangian (Eq.~\ref{eq:lagrangian_TEGR}) and the YM Lagrangian (Eq.~\ref{eq:ym_lagrangian}).

\section{\textcolor{black}{Physical bosons of the U(1, 3) adjoint representation}}
\label{sec:adjoint_representation}

The adjoint representation of a Lie group $G$ describes how the group acts on its own Lie algebra $\mathfrak{g}$ via the adjoint map $\text{adj}_g(X) = g X g^{-1}$ for $g \in G, \ X \in \mathfrak{g}$. This is useful for identifying the gauge fields (bosons) that participate in the interaction. In the case of $\text{U}(1,3)$, which consists of complex 4×4 matrices that preserve a Hermitian form of signature (1,3), there are 16 generators $\{\ell_I\}_{I=0}^{15}$ forming the Lie algebra $\mathfrak{u}(1,3)$ (see adjoint representation in Sec.~\ref{sec:u13}). It can be decomposed as $\text{U}(1,3) \cong \text{SU}(1,3) \times_{\mathbb{Z}_4} \text{U}(1)$\footnote{By taking the product $\text{SU}(1,3) \times \text{U}(1)$ and then quotienting by $\mathbb{Z}_4$, we effectively construct the group $\text{U}(1,3)$, which includes the necessary phase factors and maintains the unitary property with the correct determinant condition. This construction is a standard way to relate these groups and ensure the correct group structure and properties.} where the fibered product accounts for the overlap in center elements between $\text{SU}(1,3)$ and $\text{U}(1)$, and the Lie algebra accordingly splits as:
\begin{equation*}
\mathfrak{u}(1,3) = \mathfrak{su}(1,3) \oplus \mathfrak{u}(1).
\end{equation*}

Although $\text{U}(1,3)$ has rank 4 --matching that of the SM group $\text{SU}(3) \times \text{SU}(2) \times \text{U}(1)$-- it contains four additional generators due to its non-compact structure of internal mixing of time-like and space-like components. These extra generators expand the algebra but do not necessarily introduce new physical bosons.

These extra degrees of freedom are absorbed or reinterpreted through the color-linked embedding of weak interactions, where multiple SU(2) sectors (associated with each color) effectively reduce to a single SU(2) structure at low energy scales. This collapse is enabled by the Higgs mechanism (Sec.~\ref{sec:higgs_mechanism}), which provides mass to the overlapping directions and identifies a unified electroweak basis.

Consequently, the 16-dimensional adjoint representation of $\text{U}(1,3)$ breaks down to the 12-dimensional set of physical SM gauge bosons:
\begin{eqnarray}
\text{adj}_{\text{U}(1,3)} &\longrightarrow& \text{adj}_{\text{SU}(3)} \oplus (\text{adj}_{\text{SU}(2)} \oplus \text{adj}_{\text{U}(1)}),
\\
16 &\longrightarrow& \underbrace{8}_{\text{gluons}} + \;(\underbrace{6/3}_{W^\pm} + \;\underbrace{1}_{Z^0} + \underbrace{1}_{\text{photon}}) = 12,
\end{eqnarray}
taking into account the redundancies within the 16 gauge fields $\{A^I\}_{I=0}^{15}$ of the U(1,3) as follows:
\begin{itemize}[noitemsep, topsep=0pt]  
\item Eight gluon fields $\{G^a\}_{a=1}^8$ corresponding to eight $\text{SU}(3) \subset \text{SU}(1,3)$ purely-colored (spatial subgroup) generators, represented by Gell-Mann matrices $\lambda_a$ ($a = 1, \dots, 8$).  
\item Eight non-compact gauge fields with redundancies:
\begin{itemize}[noitemsep, topsep=0pt]
    \item One boson field $B$ associated with the $\text{U}(1)_\Gamma$ 
   generator: the identity $\Gamma_0$ (Eq.~\ref{eq:identity}). 
    \item One boson field $W^0$ linked to the $\text{U}(1)_y$ hypercharge generator $Y_w$ (Eq.~\ref{eq:hypercharge})
    \item Six gauge fields ($\{W^1_{c}\}_{c=r,g,b}$ and $\{W^2_{c}\}_{c=r,g,b}$) linked to six generators from three partial $\text{SU}(2)_w$ groups, $\{w_{1,c}^\pm, w_{2,c}^\pm\}_{c = r,g,b}$ that are elements of three color-triplet $\mathfrak{su}(2)$ algebras but only generate one effective $\text{SU}(2)$ electroweak group. 
    \item However, from these eight fields ($B$, $W^1_{c=r,g,b}$, $W^2_{c=r,g,b}$, $W^0$), only four physical bosons emerge via color-symmetry mixing (Eq.~\ref{eq:beta_mix}) and rotation by Higgs mechanism (Sec.~\ref{sec:predicted_angle}):  
     \begin{eqnarray}
     \begin{pmatrix}
         W^1 \\ W^2
     \end{pmatrix}   &=& \frac{1}{3} \sum_{c \in \mathcal{C}} 
     \begin{pmatrix}
         W^1_{c} \\ W^2_{c}
     \end{pmatrix},
         \\
          \begin{pmatrix}
         \gamma  \\  Z^0
     \end{pmatrix}   &=&  \begin{pmatrix}
         \cos\theta_W   &  \sin\theta_W  \\
         - \sin\theta_W  & \cos\theta_W  
     \end{pmatrix} 
     \begin{pmatrix}
         B \\ W^{0}
     \end{pmatrix}
     \end{eqnarray} where $\theta_W$ is the Weinberg angle (Eq.~\ref{eq:angle_W}), $\mathcal{C} = \{r,g,b\}$ is the color, and $W_\mu^0$ is the boson linked to the $\text{U}(1)_y$ hypercharge generator $Y_w$, which contributes to the effective electroweak $Z^0$ boson and the final $\gamma$ photon by mixing with $B_\mu$. The remaining four generators in $\mathfrak{u}(1,3)$ correspond to color directions that are spontaneously merged and do not manifest as independent physical gauge bosons. Their role is confined to the high-energy unifying regime, from which the SM symmetries emerge effectively.  
\end{itemize}
\end{itemize}
 
This structure ensures all SM gauge fields ($\{G^a\}_{a=1}^8$, $W^1$, $W^2$, $Z^0$, $\gamma$) transform as adjoints under their respective subgroups, while the full $\text{U}(1,3)$ adjoint representation ($\mathbf{16}$) integrates gravity via the original $\text{U}(1)_\Gamma$ sector (identity). The Higgs VEV projects out the non-SM components, leaving only the massless SM adjoints at low energies.

%\[ \delta A_\mu^a = \frac{1}{g} \partial_\mu \varphi^a + f^{abc} \varphi^b A_\mu^c, \]

\color{black}

% as detailed in Sec.~\ref{sec:higgs_mechanism}.

%The following is incomplete and incorrect:

%\textcolor{black}{The $\text{U}(1,3)$ gauge field $A_\mu = A_\mu^I \ell_I$ ($I=1,\dots,16$) can be written as:  
%\[
%A_\mu = \begin{pmatrix} 
%B_\mu + Y_\mu & W_\mu^+ & W_\mu^- & \cdots \\ 
%W_\mu^- & G_\mu^3 + \frac{1}{\sqrt{3}}G_\mu^8 & G_\mu^1 - iG_\mu^2 & \cdots \\ 
%W_\mu^+ & G_\mu^1 + iG_\mu^2 & -G_\mu^3 + \frac{1}{\sqrt{3}}G_\mu^8 & \cdots \\ 
%\vdots & \vdots & \vdots & \ddots 
%\end{pmatrix},  
%\]  
%where off-diagonal terms include lepto-color bosons (massive) and diagonal terms contain SM fields (massless).}  
% Bibliography

%% [A] Recommended: using JHEP.bst file
\bibliographystyle{JHEP}
\bibliography{biblio.bib}

%% or
%% [B] Manual formatting (see below)
%% (i) We suggest to always provide author, title and journal data or doi:
%% in short all the informations that clearly identify a document.
%% (ii) please avoid comments such as "For a review'', "For some examples",
%% "and references therein" or move them in the text. In general, please leave only references in the bibliography and move all
%% accessory text in footnotes.
%% (iii) Also, please have only one work for each \bibitem.

\end{document}